\documentclass[letterpaper,8pt]{extarticle}
\usepackage{natbib}
\setcitestyle{authoryear,open={(},close={)}}
\usepackage{local}
\usepackage[left=1in,top=.65in,bottom=.65in,right=1in]{geometry}

\title{Uncertainties in critical slowing down indicators of observation-based fingerprints of the Atlantic Overturning Circulation}

\author{%
\textbf{Maya Ben-Yami,\textcolor{Accent}{\textsuperscript{1,2*}}
Vanessa Skiba, \textcolor{Accent}{\textsuperscript{2}}
Sebastian Bathiany, \textcolor{Accent}{\textsuperscript{1,2}}
Niklas Boers,\textcolor{Accent}{\textsuperscript{1,2,3}}}\\
\begin{small}\textcolor{Accent}{\textsuperscript{1}}Earth System Modelling, School of Engineering and Design, Technical University of Munich, Munich, Germany\\ 
\textcolor{Accent}{\textsuperscript{2}}Potsdam Institute for Climate Impact Research, Potsdam, Germany\\
\textcolor{Accent}{\textsuperscript{3}}Department of Mathematics and Global Systems Institute , University of Exeter, Exeter, UK\\
\textcolor{Accent}{\textsuperscript{*}}Correspondence: \textcolor{Accent}{maya.ben-yami@tum.de} \\ \end{small}
}

\date{}

\begin{document}
\maketitle
\bigskip
\section{Abstract}
\begin{doublespacing}

Observations are increasingly used to detect critical slowing down (CSD) to measure stability changes in key Earth system components. However, most datasets have non-stationary missing-data distributions, biases and uncertainties. Here we show that, together with the pre-processing steps used to deal with them, these can bias the CSD analysis. We present an uncertainty quantification method to address such issues. We show how to propagate uncertainties provided with the datasets to the CSD analysis and develop conservative, surrogate-based significance tests on the CSD indicators. We apply our method to three observational sea-surface temperature and salinity datasets and to fingerprints of the Atlantic Meridional Overturning Circulation derived from them. We find that the properties of these datasets and especially the specific gap filling procedures can in some cases indeed cause false indication of CSD. However, CSD indicators in the North Atlantic are still present and significant when accounting for dataset uncertainties and non-stationary observational coverage.

\newpage
\section{Introduction}
\label{sec:intro}
In recent years there has been increasing focus on non-linearities and the potential of abrupt transitions in the Earth system, especially in response to anthropogenic greenhouse gas emissions. Of particular interest are systems that have multiple stable equilibrium states, and so could rapidly transition in a self-perpetuating way to a different state once a critical forcing threshold is reached \citep{Mckay2022MultiplePoints}. When such systems approach a transition to a different state in response to gradual changes in forcing, they may exhibit so-called critical slowing down (CSD), in which their response to perturbations changes in a characteristic manner \citep{Dakos2008SlowingChange}. CSD can be a sign of a forthcoming transition and may in certain situations be used to anticipate it; statistical signs of CSD, such as increasing variance or autocorrelation, have hence also been termed early-warning signals \citep{Scheffer2009Early-warningTransitions}. CSD has been identified in observations of numerous Earth system components that have been identified as tipping elements \citep{McKay2022ExceedingPoints}. These include the the Greenland Ice Sheet \citep{Boers2021CriticalPoint}, the Atlantic Meridional Overturning Circulation (AMOC) \citep{Boers2021Observation-basedCirculation, Michel2022EarlyReconstruction}, the Amazon rainforest, \citep{Boulton2022Pronounced2000s} as well as other parts of global vegetation \citep{Smith2022ReliabilitySeries, Smith2022EmpiricalResilience}.

However, CSD indicators such as the variance and lag-one autocorrelation \citep{Scheffer2009Early-warningTransitions} or the restoring rate \citep{Boers2021Observation-basedCirculation, Held2004DetectionFingerprinting} are calculated from observational datasets that are not optimised to capture higher-order statistics. Observational datasets employ a variety of methods to combine data from different instruments, adjust observational biases and fill in missing grid cells. These methods are tuned to best capture mean global trends in the data, sometimes at the cost of underlying statistical properties. For example, variance may decrease just as a result of increasing data coverage. This can be caused by an increasing numbers of observations that can be taken into account for each reported value, e.g. by taking the mean over samples of increasing size and correspondingly reduced standard error. However, this is just a simplified example, and as each observational dataset has its own specific methods of data assimilation and infilling, it is not possible to generalize the effect that observational dataset uncertainties would have on higher-order statistics. For an in-depth investigation of data aggregation effects using remote sensing data we refer to \citet{Smith2022ReliabilitySeries}. 
In this work we focus on sea-surface temperature (SST) and salinity based CSD indicators for the AMOC. We show how dataset uncertainties can be incorporated into the CSD analysis, and how the standard significance testing methods can be modified to account for the influence of different infilling methods. 

This study is based on the work by \cite{Boers2021Observation-basedCirculation} (hereafter B21). B21 analysed SST- and salinity-based proxies of the AMOC strength \citep{Rahmstorf2015ExceptionalCirculation, Caesar2018ObservedCirculation} to investigate whether a declining stability of the AMOC can be detected from statistical indicators. The AMOC is a key element of the Earth's climate system, transporting large amounts of heat and salt northward in the upper layers of the Atlantic Ocean. Paleoclimate proxy evidence as well as theoretical considerations suggest that the AMOC is bistable, with a second, substantially weaker circulation mode in addition to the present strong mode \citep{Henry2016NorthGlaciation,Stommel1961ThermohalineFlow, Rahmstorf2002OceanYears, Bohm2015StrongCycle, Broecker1985DoesOperation, Ganopolski2001RapidModel}. The bistability of the AMOC has recently been supported by comprehensive high-resolution model simulations \citep{Jackson2018HysteresisGCM}. There are several lines of proxy- and observation-based evidence suggesting that the AMOC has indeed  weakened in the last decades to centuries \citep{Caesar2021CurrentMillennium}, although the decline and its cause are still controversial \citep{Latif2014Natural1900, Kilbourne2022AtlanticUncertain}. Finally, comprehensive models predict that the AMOC will weaken further under anthropogenic global warming \citep{IPCC_2021_WGI_Ch_4}.

The most commonly used CSD indicators are an increase in the variance and autocorrelation of a timeseries. However, these indicators can result in false positives, as an increase in the variance or autocorrelation can also be caused by a corresponding statisical change in the external conditions, often modeled as noise. To avoid such false positives, B21 introduced the restoring rate $\lambda$, estimated under the assumption of non-stationary correlated noise driving the system. For a system in state $x$ close to equilibrium, we can linearize about the equilibrium state and thus the dynamics can be approximated as $\frac{dx}{dt}\approx\lambda x+ \eta$, where $\eta$ stands for random external perturbations. $\lambda$ can thus be estimated by regressing and estimating of the derivative $dx/dt$ against $x$. One can then avoid false CSD indicators caused by the properties of $\eta$ by performing this regression with a generalized least square algorithm that assumes noise with varying autocorrelation (for more details see B21). $\lambda$ is negative for systems close to a stable state, and when a multistable system approaches a critical transition, $\lambda$ increases to 0 from below. We focus on $\lambda$ in the main text of the paper; corresponding analyses and figures for the variance and autocorrelation can be found in the supplementary materials. 

A prerequisite for a statistically significant increase in CSD indicators is a sufficiently long time series. Direct observations of the AMOC strength in the Northern Atlantic only go back to 2004 \citep{RAPID}. Consequently, numerous AMOC fingerprints based on observations spanning longer time periods have been suggested, which are thought to reflect variations in the strength of the AMOC. As the AMOC transports heat and salt northward, SSTs and salinity profiles are commonly used as AMOC fingerprints. B21 took two approaches to identifying CSD indicators for the AMOC. The first is to look for CSD indicators in previously identified fingerprints that are constructed by averaging SST or salinity over a specific region \citep{Chen2018GlobalCirculation, Zhang2017GeophysicalLetters, Rahmstorf2015ExceptionalCirculation, Caesar2018ObservedCirculation}. For example, one such fingerprint was proposed by \cite{Caesar2018ObservedCirculation} and is calculated by taking the average SST in the subpolar gyre region minus the average global SST (see also \cite{Rahmstorf2015ExceptionalCirculation}). The second approach is to calculate the CSD indicators for each grid cell in the SST or salinity dataset, and look at the regions that are thought to be related to the AMOC strength. For example, if the AMOC weakens, salinity is accumulated along its main transport path, and thus the changes in near-surface salinity along the Gulf Stream and North Atlantic Current are thought to reflect changes in the strength of the AMOC \citep{Rahmstorf2002OceanYears}. B21 found significant increases in $\lambda$ both in the SST and salinity fingerprints and on spatially explicit maps, and both of these approaches will be used in this work.

B21 used three observational datasets: the HadISST1 \citep{Rayner2003GlobalCentury} and ERSST \citep{ERSST} datasets for SSTs and the EN4.2.1 dataset \citep{Good2013EN4:Estimates} for salinity profiles. They provide smooth global fields from 1871 to present for the SST data and from 1900 to present for the salinity data. 
In this work we will use an updated version of the EN4 salinity data (EN4) and for further robustness testing additionally use the HadSST4 \citep{Kennedy2019AnSet} and HadCRUT5 \citep{Morice2021AnSet} SST datasets, which date back to 1850. 

Although this study will focus on SST and salinity datasets and on CSD indicators, the work presented here can be generalised to many other datasets and higher-order statistics (for example, \cite{Shi2022GlobalCentury}).

\begin{figure}[htb!]
    \centering
    \includegraphics[width=\textwidth]{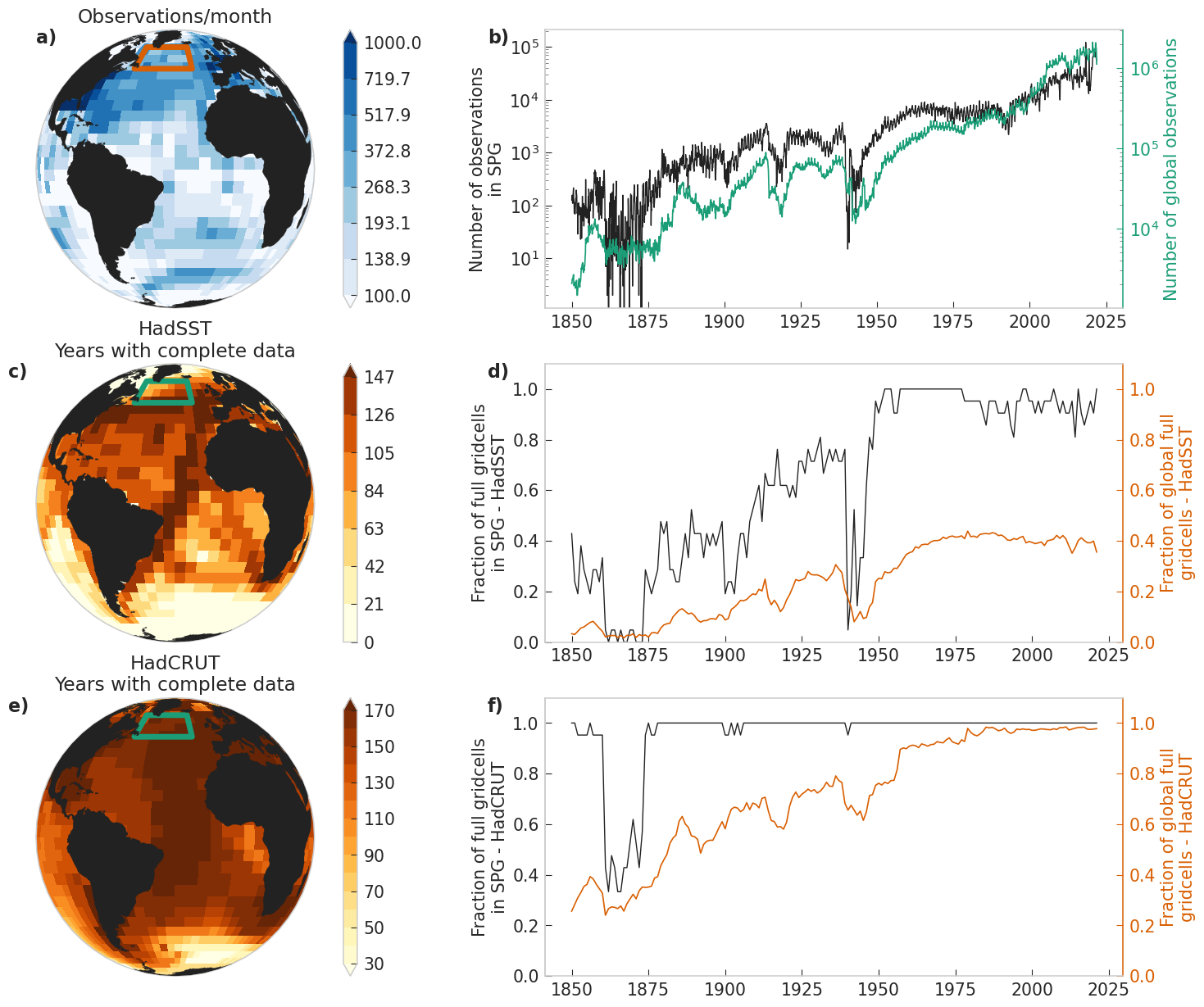}
    \caption{a. Mean number of SST observations per month in HadSST and HadCRUT for the time span 1850-2021 in each grid cell. b. Number of observations per month in HadSST and HadCRUT  in the subploar gyre (SPG, turqoise) and globally (black). Raw observations are the same for both HadSST and HadCRUT. c. Number of years with complete annual mean data (see Methods) for HadSST. d. Fraction of full grid cells after taking the annual mean in the SPG (orange) and globally (black) in HadSST. e. Same as (c) but for HadCRUT. f. Same as (d) but for HadCRUT. The SPG area is shown as a square on the maps.}
    \label{fig:nobs}
\end{figure}

\section{Uncertainty ensembles}
\label{sec:uncertainty}
\begin{figure}[htb!]
    \centering
    \includegraphics[width=\textwidth]{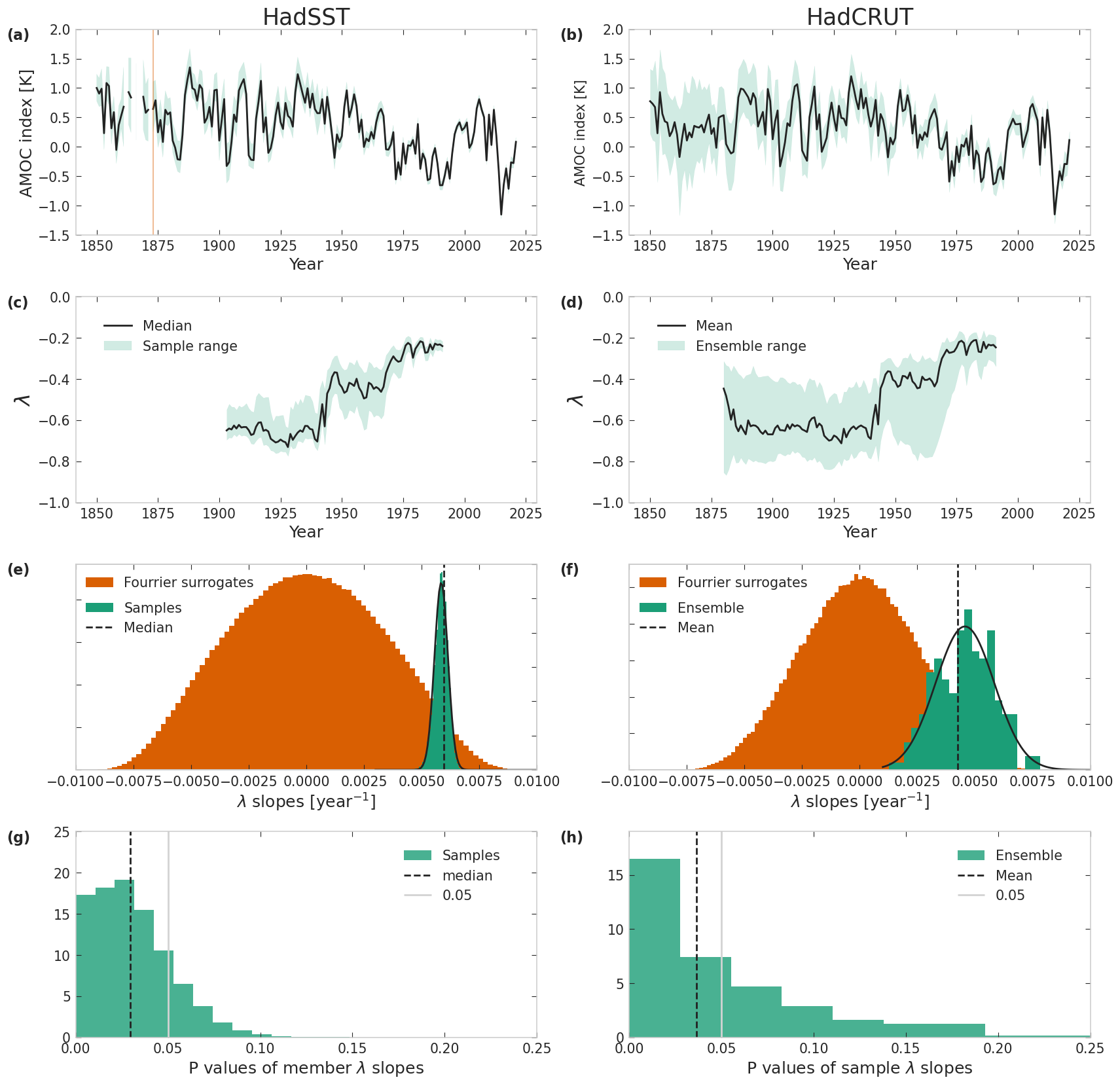}
    \caption{Significance of trend in $\lambda$ of subpolar gyre AMOC index in HadSST (a,c,e,g) and HadCRUT (b,d,f,h). a,b. Median (mean) AMOC index (black) and min-max range (turqouise) of 20000 samples (200 ensemble members) for HadSST (HadCRUT). We use the mean instead of the median for HadCRUT as that is the default product a user would download when not investigating uncertainties. c,d. Same as (a,b) but for the $\lambda$ of the AMOC indices, computed using a  window size of 60 years. e. Distribution (orange) of linear trends of $\lambda$ computed from 100 Fourier surrogates for each AMOC index sample from HadSST; distribution (turquoise) of linear trends of $\lambda$ of samples from HadSST with a fitted gausian distribution (solid black). The linear trend of the median HadSST index is shown in dashed black. f. Same as e but for HadCRUT, using fourier surrogates of the ensemble member AMOC indices, and the linear trend of the mean. g. p-value of linear trend of $\lambda$ of each sample of the HadSST AMOC index with respect to its own 100 Fourier surrogates. The p-values of the median with respect to 10000 Fourier surrogates is shown as a dashed black line, and the 0.05 significance value is shown as a solid grey line. f. same as g but for the AMOC index of the HadCRUT ensemble members, with the p-value of the mean AMOC index as a dashed black line.}
    \label{fig:indices}
\end{figure}

In this section, we asses how uncertainties provided with the observational datasets propagate to uncertainties in the CSD indicators (or other higher-order statistics). The only uncertainty provided for EN4.2.2 is the uncertainty associated with the analysis method, and that estimate has issues that limit its usefulness for our analysis; for example, in some areas more observations actually increase the analysis uncertainty (see \cite{Good2013EN4:Estimates}). Thus, in this section, we make use only of the uncertainties provided with the HadCRUT5 and HadSST4 datasets. See Methods for a detailed discussion of the analysis and processing procedures of these datasets, as well as of the provided uncertainties.

We first use both datasets to calculate the AMOC fingerprint proposed by \cite{Caesar2018ObservedCirculation} (Fig \ref{fig:indices}). The advantage to using both datasets is that they represent two different ways of dealing with missing data. HadSST simply has no data where there are no observations. In the HadCRUT infilled dataset those data points are filled in, with the exception of a few remaining gaps. Thus, HadSST only has continuous coverage of the SPG from 1873, after which the number of full grid cells gradually increases (Fig \ref{fig:nobs}c), whilst HadCRUT has about the same number of grid cells in the SPG over the whole period (Fig \ref{fig:nobs}e). Thus, in HadSST the AMOC index could be biased towards a sub-region of the SPG with more observations in certain time periods. The increasing number of averaged full grid cells also affects the variance, causing a decrease of variance with time that is unrelated to the underlying properties of the ocean. On the other hand, infilling the missing grid cells in HadCRUT comes with the uncertainties of the infilling method. 
\begin{figure}[htb!]
    \centering
    \includegraphics[width=\textwidth]{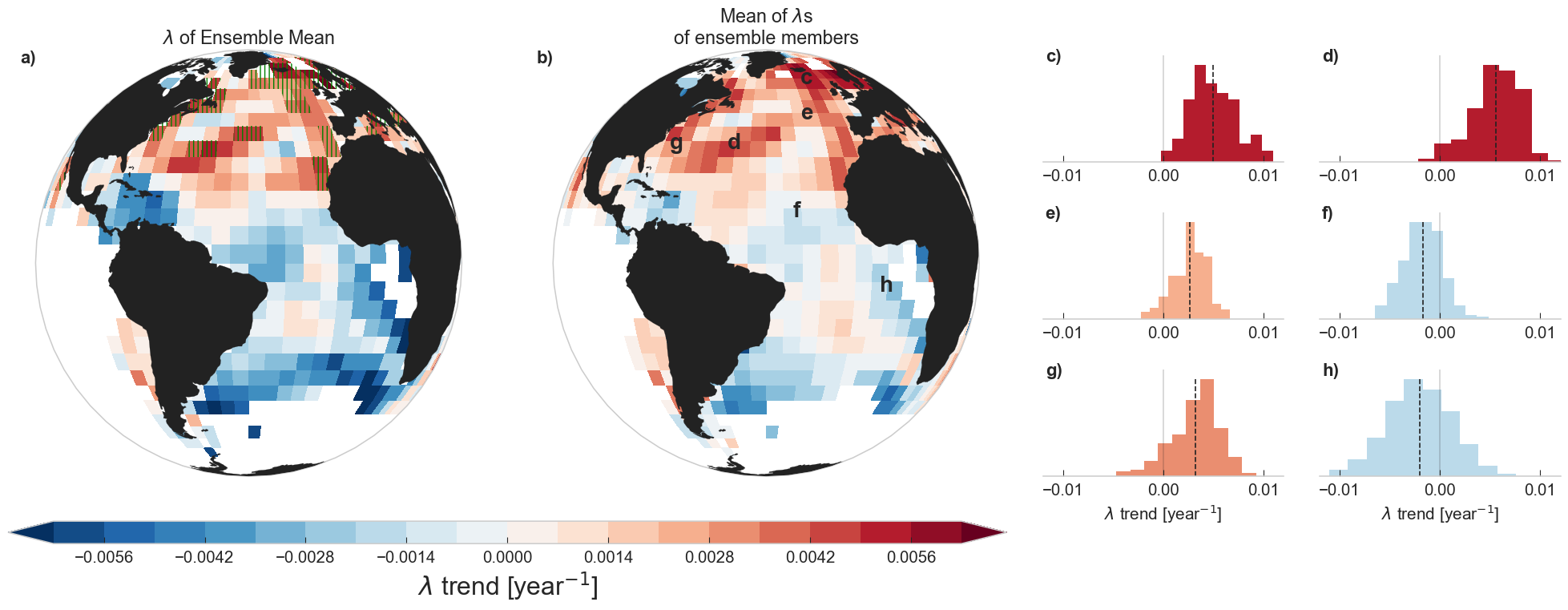}
    \caption{a. Linear trends of $\lambda$ time series computed from the ensemble mean in the HadCRUT dataset. b. Mean of the linear trends computed individually from each $\lambda$ time series of the 200 ensemble members. c-h. Distributions of 200 trends for locations marked with corresponding letters in (b), with a vertical black dashed line showing the ensemble mean, and light solid grey lines at 0. The green vertical hatches in (a) indicate the regions where $\mu-2\sigma>0$ for the Gaussian fitted to the ensemble distribution, i.e. where most of the uncertainty ensemble trends are positive. 
    }
    \label{fig:hadcrut}
\end{figure}

For HadSST, we create an ensemble that captures the uncertainty at each grid cell by sampling the measurement and sampling uncertainty around each bias ensemble member, creating 200x100 samples. For HadCRUT, we simply make use of the provided 200-member uncertainty ensemble, which already accounts for all aspects of uncertainty, including the interpolation uncertainty. In both cases, we calculate the AMOC index following \citet{Caesar2018ObservedCirculation} for each sample or member and then calculate its $\lambda$ (Fig \ref{fig:indices}c,d). Because of the infilling of missing values, the HadCRUT index and $\lambda$ have a larger uncertainty range than HadSST, especially in the early years. This is also reflected in the distribution of trends: the fitted Gaussian of $\lambda$ trends of the HadSST samples is narrow ($\mu=0.0059,\sigma=0.0003$), whilst the one for HadCRUT is wider ($\mu=0.0046,\sigma=0.0013$)(Fig \ref{fig:indices}e,f). We choose to show the median and mean time series for HadSST and HadCRUT, respectively, as those are the time series a user would obtain when downloading the data from the Met Office website (www.metoffice.gov.uk/hadobs/hadcrut5/, www.metoffice.gov.uk/hadobs/hadsst4/). In general, the median is a better choice, as taking the mean changes the statistical properties. 

However, the magnitude of the trend is of lesser interest to us than whether or not it is statistically significant. To test for significance, we calculate 1000 Fourier surrogates (see Methods) from each AMOC time series, and use the obtained linear trends from the $\lambda$ time series for each sample or member individually to calculate a p-value (Fig \ref{fig:indices}g,h). 85.56\% of HadSST and 65\% of HadCRUT p-values are below 0.05, and 99.7\% and 84.5\% are below 0.1, showing that even considering the dataset uncertainties, the increase in $\lambda$ of the SPG-based AMOC index is significant.

For the HadCRUT dataset, we can take the analysis a step further - most regions of the Atlantic have enough full values (Fig \ref{fig:nobs}e) to compute $\lambda$ for each grid cell, and repeat this for all ensemble members. The resulting map shows a significant positive linear trend of $\lambda$ in the North Atlantic (Fig \ref{fig:hadcrut}a), similar to that seen in B21 for the HadISST dataset (\cite{Boers2021Observation-basedCirculation}). The trends in the ensemble members vary, but their mean is very similar to the trend of the ensemble mean (Fig \ref{fig:hadcrut}b,c-h). We also fit a gaussian to the ensemble distribution at each gridpoint and show the regions where most of the uncertainty ensemble $\lambda$ trends are positive ($\mu-2\sigma>0$). These regions are along the North Atlantic Current and in the Greenland, Iceland and Norwegian Seas (Fig \ref{fig:hadcrut}a). 

Figures \ref{fig:ar1_indices}, \ref{fig:std_indices}, \ref{fig:HadCRUT_ar1} and \ref{fig:HadCRUT_std} show the same calculations for the variance and autocorrelation. The autocorrelation behaves similarly to $\lambda$, although with a less significant increase. The variance has no overall trend, due to a marked decrease from 1850 to the 1950s, which is likely caused by the increasing number of observations for the reasons explained above.

\section{Global significance estimation with surrogates}
\label{sec:surrogates}

In addition to an uncertainty estimation, it is also important to calculate the statistical significance of our higher-order statistic of interest. When we want to test a statistic of a time series $x_t$, for example the linear trend of its lambda time series, denoted by $s_{\lambda}(x_t)$, this is generally done by generating surrogate time series of $x_t$, $\bar{x}^i_t$ \citep{Theiler1992TestingData}. The values of the statistic for many surrogate time series $s_{\lambda}(\bar{x}^i_t)$ can then be used as a distribution to estimate the significance of the actual $s_{\lambda}(x_t)$. Our null model determines the properties that the surrogate time series need to have. When calculating CSD indicators, our null model should be a time series that has the same autocorrelation structure and variance as $x_t$, but is otherwise random. Such a time series can be produced either by measuring a finite number of autocorrelation properties of $x_t$ and generating an according time series, or by the method of Fourier surrogates, where the Fourier phases are randomly shuffled (see Methods). These surrogates can be modified to include the effects that interpolation methods or lack of data have on the time series (e.g. removing data points that are missing in the original time series). 

It is important to note that in the case of calculating CSD indicators, our conservative null model is an SST or salinity time series with given properties, not a $\lambda$ time series. In this regard, the surrogate analysis of this work is more conservative than that in B21, who considered surrogates of the $\lambda$ time series. This is because the autocorrelation structure of $\lambda_t$ has a non-trivial dependence on the autocorrelation of $x_t$. By generating surrogates of $\lambda_t$ as in B21 one ignores the wider range of autocorrelations that $\lambda_t$ can have given the autocorrelation of $x_t$. This can result in a narrower distribution of surrogate $\lambda$ trends, and the significance of $s_{\lambda}(x_t)$ is thus overestimated. It is therefore important to generate time series from $x_t$ and not $\lambda_t$, even if the latter is computationally more costly. Note that this also implies that Fourier surrogates are not generally suited to test trend significance in arbitrary time-correlated time series. They should only be used in situations where the trend of a sliding-window higher-order statistic is estimated from a given time series, and one has access to that time series to compute surrogates from. 

\begin{figure}[htb!]
    \centering
    \includegraphics[width=\textwidth]{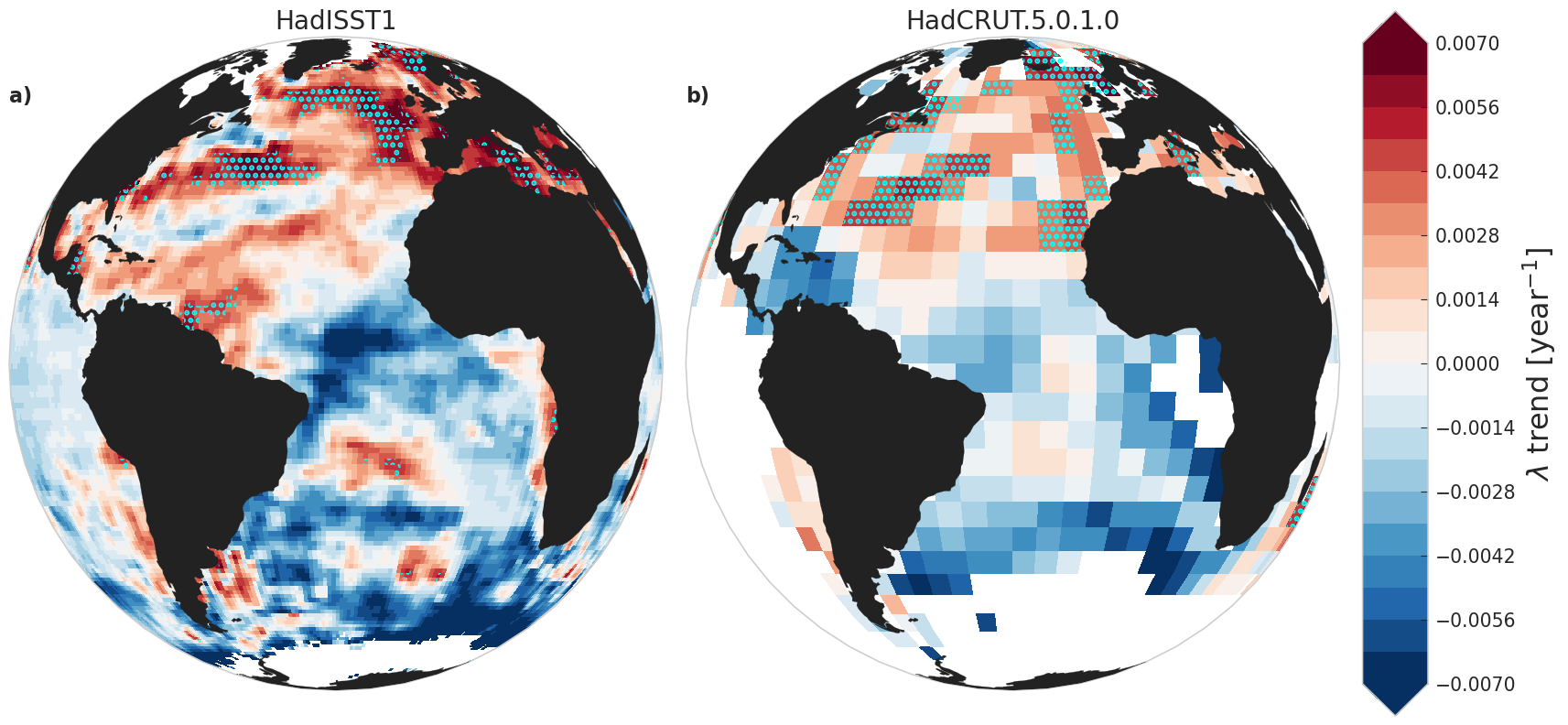}
    \caption{a. Linear trends of $\lambda$ time series for the HadISST1 data. b. Same as (a) but for the HadCRUT mean dataset. Light blue stippling shows the regions where the trends are significant at 95th percentile, calculated from 1000 Fourier surrogates for each cell for HadISST (a) and 1000 AR(2) surrogates for each cell for HadCRUT (b). See Methods for more details.}
    \label{fig:sstsig}
\end{figure}

The HadISST dataset is globally interpolated, and thus the time series at each gridpoint is complete, and we can use Fourier surrogates to calculate its regions of significance (Fig \ref{fig:sstsig}a). Due to the holes in the HadCRUT dataset we cannot use Fourier surrogates (see Methods) and thus use AR(2) surrogates instead to calculate the significant regions (Fig \ref{fig:sstsig}b).

For both datasets, the region of increasing $\lambda$ extends over the whole North Atlantic, but only a smaller region shows statistically significant change at the 0.05 confidence level. For HadISST, significant increase only occurs along the North Atlantic Current, the density-driven part of the AMOC, whilst for HadCRUT the significant increase occurs at the northern edge of the sub-tropical gyre, along the North Atlantic Current and in the eastern sub-polar North Atlantic, including the Greenland, Iceland and Norwegian seas.

\begin{figure}[htb!]
    \centering
    \includegraphics[width=\textwidth]{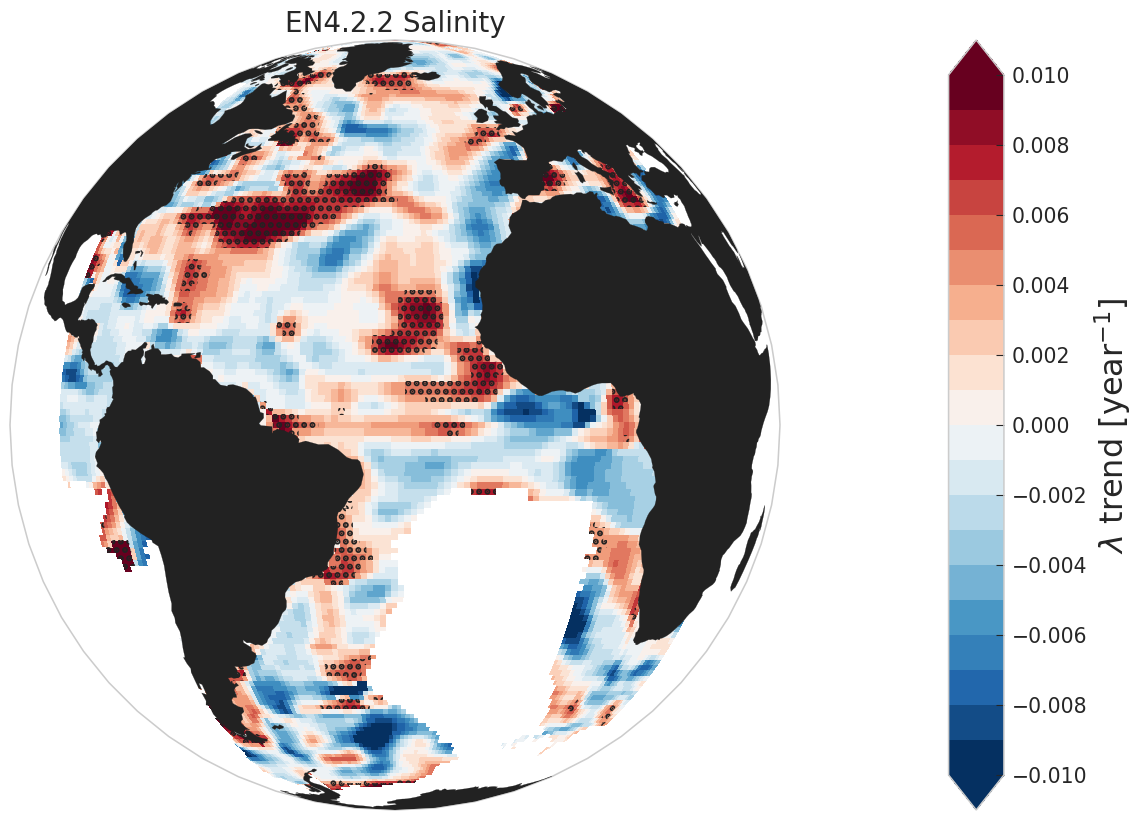}
    \caption{a. Linear trend of $\lambda$ time series for the average top 300m salinity in the EN4 dataset. Black stippling shows the regions of 95th percentile significance calculated using 1000 AR(2) surrogates modified following the EN4 analysis method with an observational weight bound of 0.5. These significance regions are indistinguishable from those calculated from the unmodified surrogates (see Figure \ref{fig:elim_options}). Only regions between 90$^\circ$W and 30$^\circ$E with 60+ years of observational weight above 0.05 are shown. See Figure \ref{fig:w_years} and Methods for details.}
    \label{fig:salsig}
\end{figure}

As opposed to the SST datasets, the infilling method in EN4 uses information from previous times, as well as a climatology. This has the potential to cause false indications of CSD (see Methods for a full discussion). Thus when generating surrogates for a time series from the EN4 salinity dataset, we must consider the effect the lack of data and analysis method has on the earlier years. The full analysis process is too complex to reproduce when generating the surrogates. However, we can reproduce the specific effect that the analysis procedure has on the calculated CSD indicators. To do this, we use the observational weights provided with the dataset (see e.g. Fig \ref{fig:nprof}h-m). These observational weights were produced by setting all observational values to one and the climatology to 0 and rerunning the analysis that produced the infilled dataset \citep{Good2013EN4:Estimates}. The resulting weight $w$ represents the amount of information in the given analysis value that comes from observations. It should be noted that this observational information comes from the whole globe, and so $w$ can be high even if there is no observation at the specific gridpoint. However, a low $w$ value is still a good indicator of a datapoint where the persistence based forecast dominates. 

For each time series, we use the autocorrelation properties of its last 40 years to generate AR(2) surrogates. Then, for each month that has an observational weight below some limit $w_0$, we replace the surrogate value with the persistence-based forecast (Eq \ref{eq:forecast}). This creates the same spikes in the surrogates that are present in the analysis data, and thus modifies the autocorrelation structure in a similar way (Fig.~\ref{fig:modification}). 

Using the unmodified surrogates, there are regions of significantly positive linear trends of $\lambda$ for the EN4.2.2 salinity data at the northern edge of the sub-tropical gyre and along the North Atlantic Current (Fig \ref{fig:elim_options}e). The modification of the surrogates to account for the non-stationary data coverage results in spurious increasing and decreasing $\lambda$ trends (Fig \ref{fig:elim_options}b-d), with slightly positive trends in the North Atlantic and negative ones near the equator, as would be expected from the autocorrelation values of those regions. However, the magnitude of these false trends is much smaller than the increases seen in the analysis data in some regions, and thus the regions of significance in the Atlantic Ocean remain basically unchanged when we use the modified surrogates (Fig \ref{fig:salsig}). This is true regardless of the limit value chosen for $w_0$ (Fig \ref{fig:elim_options}f-h).
\begin{figure}[htb!]
    \centering
    \includegraphics[width=\textwidth]{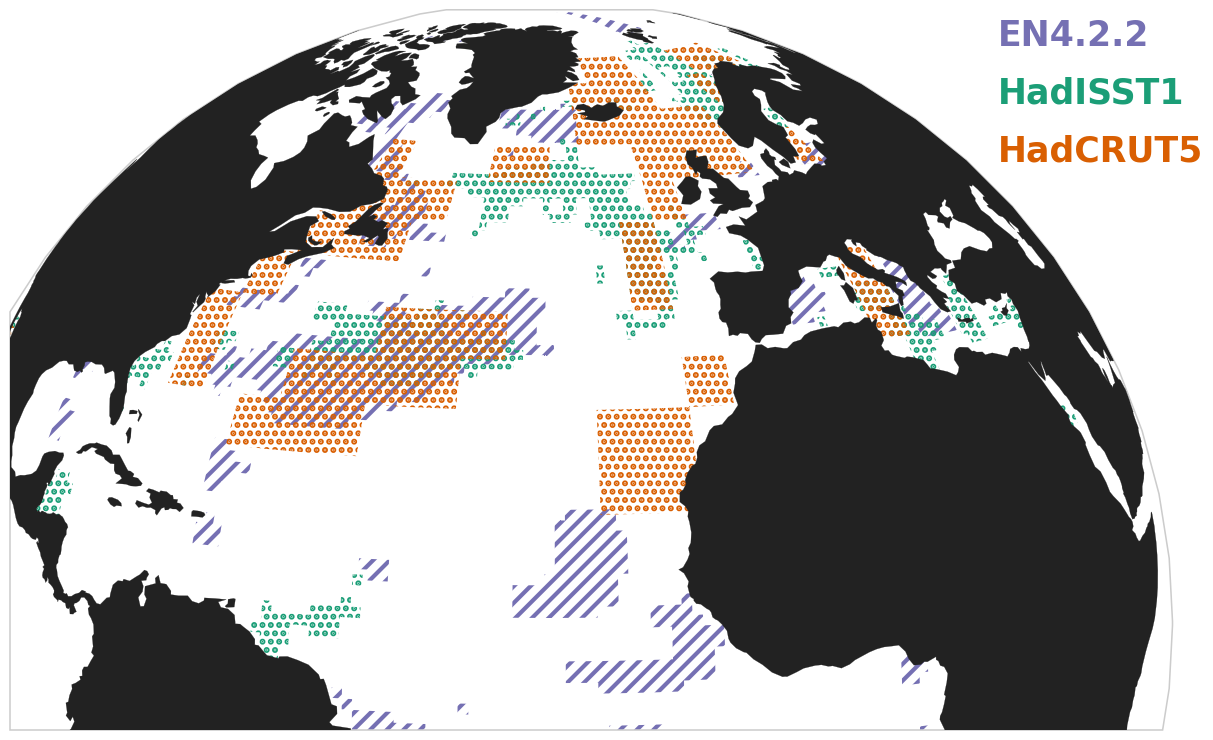}
    \caption{Regions in the Northern Atlantic with a statistically significant increase in $\lambda$ for the EN4 salinity (purple), HadISST SSTs (turquoise) or HadCRUT SSTs (orange). Significance is calculated from 1000 surrogates for each grid cell (see text and Methods).}
    \label{fig:sig_regions}
\end{figure}
The global surrogate analyses for variance and autocorrelation can be found in Figs.~ \ref{fig:SST_significance_ar1}, \ref{fig:SST_significance_std}, \ref{fig:salinity_significance_ar1} and \ref{fig:salinity_significance_std}. The results for the autocorrelation are again similar to the results for $\lambda$ estimated under the assumption of non-stationary correlated noise. But although the regions of positive trends in the North Atlantic are spatially coherent with those of $\lambda$, the regions with significant (p<0.05) increase are instead at the northern Gulf Stream and its extension into the Atlantic Ocean (Fig \ref{fig:ar1_sig_regions} and \ref{fig:both_sig_regions}). The variance of each dataset has a different result for the trends and their areas of significance. In contrast to the autocorrelation and $\lambda$, the surrogates calculation for the variance in EN4 (Fig.~\ref{fig:salinity_significance_std}) is a clear example of the utility of modifying the AR2 surrogates to match the analysis method. Without modification the whole of the Atlantic seems to have a significant increase in variance. But once the effect of the analysis method is incorporated, no significant regions remain, and we recognise the increase in variance as spurious, caused by the analysis procedure of EN4.

\section{Discussion and conclusions}
We have addressed a common problem that arises when CSD indicators are computed from pre-processed observational data, namely, that the observational datasets have inherent and potentially non-stationary uncertainties and biases that could influence the analysis \citep{Smith2022ReliabilitySeries}. As well as complementing the analysis of B21, this work can thus be used as a basis for observational uncertainty analysis of other higher-order statistics.

Using the uncertainties provided with the datasets, we estimate an uncertainty range of the linear trend of $\lambda$ in the SST-based AMOC index (Fig \ref{fig:indices}) of 0.0059$\pm$0.0003 for the HadSST4 and 0.0046$\pm$0.0013 for the HadCRUT5 dataset, respectively. We have also updated the surrogate significance analysis of B21 for the global SST and salinity data (HadCRUT5, HadISST1 and EN4.2.2), and find that our more conservative significance test reduces the area of significant CSD indicators compared to that in B21. However, the three datasets show significant $\lambda$ increases in the area of the North Atlantic Current and in the Greenland, Iceland and Norwegian Seas (Fig \ref{fig:sig_regions}).

Whilst our work demonstrates that the indication of CSD in the chosen AMOC fingerprints is not due to any inherent properties of the observational datasets, it is beyond the scope of this work to investigate whether or not this is a clear indication that the AMOC is destabilizing. Studies describing AMOC fingerprints such as the subpolar gyre SST index primarily focus on the agreement of multidecadal trends in the streamfunction strength of the AMOC and that of the fingerprint \citep{Caesar2018ObservedCirculation}. In fact, in global climate models the interannual variablity of these fingerprints is not strongly correlated with the streamfunction strength at 26.5N (see e.g. Fig 6c of B21). On the other hand, the AMOC is a complex three dimensional system, and we would not expect the variability of the subpolar part of the AMOC to match that at subtropical latitudes \citep{Jackson2022The1980}. However, there are yet to be any studies showing that signs of destabilization along the North Atlantic Current or in the SPG are in fact signs that the large-scale AMOC is destabilizing. In fact, the North Atlantic SPG has recently been identified as a separate tipping element \citep{Mckay2022MultiplePoints, Swingedouw2021OnModels, Sgubin2017AbruptModels}, and thus some of the identified CSD indicators may be caused by only this subset of the Atlantic Ocean approaching a transition. It is also still debated whether the SPG SSTs truly reflect AMOC variability and -- if yes -- whether that relation will hold under future climate change \citep{Chafik2022IrmingerVariability, He2022ACirculation, Li2022Century-longExplanation, GhoshTwoWarming}. 

However, the fact that the areas of significant CSD indicators on the map are much smaller in our analysis than in B21, does strengthen the case for an AMOC destabilization. In B21, the whole North Atlantic was marked as significant, as well as large regions in the South Atlantic. This is not a result one would expect if the CSD was caused by a weakening AMOC. In this work, the significance is reduced to smaller regions in the North Atlantic that are more typically associated with the path of the warm branch of the AMOC: the North Atlantic Current, the Irminger Sea and the Greenland, Iceland and Norwegian Seas. These CSD indicators could be a sign of AMOC destabilization, as the SST and salinity in these regions would be sensitive to the strength of the AMOC \citep{Rahmstorf2002OceanYears}. The Irminger and Iceland basins in particular have been recently identified as the centres for subpolar AMOC variability, as opposed to the subpolar gyre \citep{Lozier2019AAtlantic, Chafik2022IrmingerVariability}. Finally, note that the definition of significance in this work is conservative. Significance is calculated on a point-by-point basis and not globally, and we do not take the spatial coherence of positive trends in the North Atlantic into account in our significance testing. 

Together with the computational expense, the uncertainty estimates provided with the observational datasets are the most important factors for a reliable uncertainty estimation of CSD. Although running the full analysis algorithm that is used to create the observational datasets is beyond the scope of this work, we are able to make modifications to the surrogates that influence the statistical properties of the data in a similar way to the full analysis. However, we cannot estimate the complete uncertainty on the CSD indicators of the salinity dataset EN4.2.2 because the analysis but not the observational uncertainties are provided. Even so, our results show that the influence of observational analysis methods on higher orders statistics should be taken into account alongside the effect on the long-term mean properties.

In summary, for CSD indicators computed from observation-based SST and salinity data we have presented a comprehensive uncertainty estimation and propagation together with significance testing. Such an analysis is a prerequisite to robustly assessing the destabilization of a system from observational data. We find that data processing methods can lead to false detection of CSD (see also \cite{Smith2022ReliabilitySeries}). However, we demonstrate that such obstacles can be overcome by incorporating the data processing effects into uncertainty estimates and significance testing.

\end{doublespacing}

\bigskip
\section{Code and data availability}
The HadISST1, HadSST4, HadCRUT5 and EN4.2.2 datasets are all available at https://www.metoffice.gov.uk/hadobs/. All code used to analyse the data and generate figures will be uploaded at https://github.com/mayaby. 

\bigskip
\section{Author contributions}
MBY, VS and NB conceived the study and designed it with contributions from SB. MBY carried out the analysis. All authors discussed results, and MBY wrote the paper with contributions from all authors.

\bigskip
\section{Competing interests}
The authors declare that there are no competing interests.

\bigskip
\section{Acknowledgments}
MBY and NB acknowledge funding by the European Union’s Horizon 2020 research and innovation programme under the Marie
Sklodowska-Curie grant agreement No.956170. NB and SB acknowledge funding by the Volkswagen foundation. This is TiPES contribution \#X; the TiPES (`Tipping Points in the Earth System') project has received funding from the European Union's Horizon 2020 research and innovation programme under grant agreement No. 820970. NB acknowledges further funding by the German Federal Ministry of Education and Research under grant No. 01LS2001A. We thank John Kennedy for fruitful discussions. 

\newpage
\begin{doublespacing}

\section*{Methods}
\subsection{Analysis and processing procedures of the different datasets}

\subsubsection{HadISST}
The HadISST dataset is based on the Met Office Marine Data Bank as well as the Comprehensive Ocean-Atmosphere Data Set (COADS) \citep{Woodruff2011ICOADSArchive}, and has been temporally and spatially homogenized using an empirical orthogonal function method called reduced-space optimal interpolation. The data is bias adjusted before applying the interpolation, and the final product is a blend of the interpolated data with the original data, once that has been quality-controlled to homogenize its grid-scale variance. This infilling makes the dataset ideal for forcing atmospheric models and for comparison with coupled climate model outputs. However, the interpolation has limitations in regions with scarce observations, and so is not optimal for statistical analyses of climate variability. In addition, HadISST does not have an uncertainty estimation of either the bias adjustement, analysis method, measurement or sampling uncertainty. This makes it difficult to estimate possible effects on the CSD analysis \citep{Rayner2003GlobalCentury}.

\subsubsection{HadSST4 and HadCRUT5}
Given the lacking uncertainty information for HadISST we focus primarily on two similar SST datasets, HadSST4 \citep{Kennedy2019AnSet} and the SST part of HadCRUT5 \citep{Morice2021AnSet}. HadSST4 is based on observations from the International Comprehensive Ocean-Atmosphere Data Set (ICOADS, \cite{Woodruff2011ICOADSArchive}) gridded to a 5$^{\circ}$ by 5$^{\circ}$ grid. Various bias adjustments are applied to the data to account for the changes in historical SST measurement techniques. The HadSST4 dataset has a 200 member ensemble which explores variations of the bias scheme parameters, and in addition comes with measurement and sampling uncertainties. This uncertainty analysis is ideal for estimating uncertainties of CSD indicators. However, as the HadSST4 dataset is non-infilled, it has large gaps where there is no data. This makes a grid cell by grid cell CSD analysis impossible, and even causes difficulties when averaging data over larger regions such as the subpolar gyre. In this work, we therefore complement our analysis with SSTs from the HadCRUT5 dataset. The SST data in HadCRUT5 is based on the HadSST4 dataset, but provides an additional, more globally complete analysis dataset. The gaps in the data are filled using a Gaussian-process-based statistical method. In regions where the local observations offer an insufficient constraint this infilled data is removed, so the final dataset still has some gaps (see Fig \ref{fig:nobs}). The dataset is comprised of 200 ensemble members, which sample the reconstruction error for the Gaussian process in addition to the bias and observational uncertainty in the data. 

The number of individual SST observations has increased approximately exponentially over the last 150 years (Fig \ref{fig:nobs}). This increase in observations affects the statistical properties of the data. Primarily, an increase in observations causes a decrease in the variance, as the data in later times is an average of more values and the variance of the mean scales as $\sim\frac{\sigma^2}{n}$, where $\sigma^2$ is the variance of the individual observations and $n$ is the number of observations. The effect this would have on the autocorrelation is more difficult to determine. The more accurate values in later times could cause a higher autocorrelation due to the improved signal-to-noise ratio of the data (see \cite{Smith2022ReliabilitySeries}), but the larger range of measurement instruments used in later times could also reduce any false contributions to the autocorrelation that are related to the instruments. In all these cases, a large part of the effect would be included in the uncertainties provided with the datasets, as they account for both sampling and measurement uncertainties. 

\subsubsection{EN4.2.2}

The statistical properties of the EN4 dataset are affected by the data analysis method in a much clearer way than the other datasets. EN4 includes both global quality-controlled ocean temperature and salinity profiles and monthly objective analyses \citep{Good2013EN4:Estimates}. The profiles are direct observations from various sources, such as the World Ocean Database \citep{WOD}. They are used to obtain the globally complete analysis by calculating an optimal fit to the good profiles and profile levels in each month, given a background (prior constraint). Good profiles and levels are those which do not fail any quality-control check. The resulting optimal interpolation equations are solved using a numerical scheme. The background used for this calculation is a damped persistence-based forecast:
\begin{equation}
    \mathbf{x}_i^{f}=\mathbf{x}_i^{c}+\alpha (\mathbf{x}_{i-1}^{a} - \mathbf{x}_{i-1}^{c}),
    \label{eq:forecast}
\end{equation}

where $\mathbf{x}_i^{f}$ denotes the damped persistence forecast for month $i$, $\mathbf{x}_i^{c}$ is the climatological mean for that month, and $\alpha = 0.9$. As we are concerned here with the statistical properties of the data, the influence of using such a persistence-based forecast as the background needs to be addressed. If there are no observations for long periods of time, the analysis will relax to the climatology for that given location. If we then have a single observation, this causes a spike in the data, which relaxes back to the climatology with a monthly lag-1 autocorrelation of 0.9. In most of the Atlantic Ocean there are very few observations before the 1950s (see Fig \ref{fig:nprof}), and so the monthly autocorrelation is artificially forced to about 0.9 at the start of the time series, which also affects the yearly autocorrelation. Depending on the true autocorrelation function of the underlying time series and depending on how much of the forecast is used, this analysis method causes a biased estimate. In particular, it could cause a false indication of CSD if the true monthly autocorrelation for the more recent decades is systematically above 0.9. 
\subsection*{AMOC indices}
The SST-based AMOC index is calculated as the mean SST in the subpolar gyre minus the mean global SST, following \citet{Caesar2018ObservedCirculation}. The subpolar gyre in this work is taken as the area between 41° and 60° N and 20° and 55° E (following \cite{Menary2020Aerosol-ForcedSimulations} for ease of calculation). This is a slightly different area than that used by B21, but makes little difference at the low 5° resolution of HadSST and HadCRUT. We also take the full year instead of the winter months, as the latter method causes no substantial difference for the change in statistical properties. 

All salinity time series and global plots in this work are for the thickness-weighted mean of the upper 19 ocean layers, corresponding to the average salinity in the top 300~m of the salinity profiles. The observational weights are similarly averaged in the top 19 layers. The uncertainty is calculated by simple uncertainty propagation: $\Delta s = \sqrt{\sum\Delta s_i^2}$, where $\{\Delta s_i\}$ are the uncertainties of each individual level. This incorrectly assumes the layer uncertainties are independent, but is acceptable here as the uncertainty is not used for any quantitative analysis, but only to display an uncertainty range in Figure \ref{fig:nprof}.

\subsection*{Gaps in the data}
Both HadSST and HadCRUT have gaps in their time series. When averaging this data for different regions, we take a conservative approach: We first spatially average the monthly resolution data and then take the yearly average of the resulting time series. When taking the yearly average we only consider a year if there is data for all 12 months, otherwise it is set to NaN. However because we make the monthly-resolution spatial average first, this approach does not ensure that each grid cell in the region has data for each month of the year. Thus the value for a given year might have more grid cells contributing to one month than another, which could affect variability. For HadCRUT we also calculate CSD indicators at each grid cell where there are less than 30 missing years out of 172. 

\subsection*{Critical slowing down indicators}
The restoring rate $\lambda$, the variance and the autocorrelation are calculated in the same manner as in B21. Each time series is first nonlinearly detrended using a running mean with a 50-year window. The edges are not removed, so the detrending method is less certain at the first and last 25 years of the time series. The CSD indicators are then calculated in 60-year running windows. Note that as in B21 the $\lambda$ plotted in this study is the numerical result of the regression of $\Delta x_i$ against $x_i$, and so is related to the analytical $\lambda'$ defined in the text by $\lambda=e^{\lambda'}-1$ (when the timestep $\Delta t = 1$). As the magnitude of $\lambda$ is immaterial in this study and we are only concerned with its increase or decrease, both definitions behave similarly and are thus interchangeable for our purposes.

\subsection*{Surrogates}
Surrogates are created from the detrended SST and salinity time series. Fourier surrogates are calculated by taking the discrete Fourier transform of the time series, multiplying by random phases and then taking the inverse Fourier transform. By the Wiener-Khinchin theorem, the variance and autocorrelation function of wide-sense-stationary random processes are specified by the squared amplitudes of the (discrete) Fourier transform. Thus the Fourier surrogates preserve the variance and autocorrelation function of the original time series.

However, in this work we know the timeseries we are dealing with have been modified by the analysis process and lack of observations. When the analysis method modifies the autocorrelation, as in the case of the salinity data, the Fourier surrogates of the full time series are not a correct null hypothesis for CSD analysis, because the autocorrelation function will include information from the earlier, modified times. In addition, Fourier surrogates can only be calculated for timeseries with no missing values, and thus cannot be used for the HadCRUT global analysis. Thus, in these cases, we do not use Fourier surrogates.

\subsubsection*{AR(2) surrogates}
For the cases where Fourier surrogates cannot be computed, we choose to use AR(2) surrogates on a monthly resolution:
\begin{equation*}
    x_{t}= a_1x_{t-1}+a_2x_{t-2}+\epsilon_t\,,
\end{equation*}
where the timeseries value at time $t$, $x_t$, is determined by the value at times $t-1,t-2$ with autocorrelation coefficients $a_1,a_2$, and $\epsilon_t$ is white noise. Even though the monthly time series in the datasets are close to being AR(1) processes, the higher coefficient is needed to get the correct lag-one autocorrelation coefficient for the yearly averaged timeseries. Since the yearly average is taken before calculating CSD indicators, using an AR(2) process instead of AR(3)-AR(12) does not make a big difference as long as the annual lag-one autocorrelation is correct. 
If the estimate of monthly lag-one autocorrelation is $A_m$, the coefficients are related by:
\begin{equation*}
    a_1 = A_m(1-a_2)
\end{equation*}
We get the values of $a_1,a_2$ for each time series by calculating the true lag-one autocorrelation estimate for the last 40 year of the annual and monthly time series, $A_y, A_m$, and then calculating the $a_2$ value that minimizes the difference of the estimated $A_y', A_m'$ of the AR(2) time series from the true values. 

\subsubsection*{Modification for salinity surrogates}
It is only possible to produce perfect surrogates for the salinity analysis data by repeating the complete analysis used to create the dataset. This is not feasible for this study. We can, however, replicate the effect of the analysis method that would potentially cause spurious CSD detection, namely the relaxation back to the climatology that occurs when there is a lack of observational data. For this we utilize the observational weights provided as part of the EN4.2.2 dataset. We start with an AR(2)-based surrogate dataset of the global monthly data averaged over the levels in the top 300m. For each grid cell, we take the months that have an observational weight that is below some limit $w_0$ and replace the surrogate value with the persistence-based forecast (see equation \ref{eq:forecast}). As we did not have access to the climatology used in the EN4 analysis, we took the first year of analysis data in each grid cell as the climatology. This will in most cases be the true climatology, as very few cells have observational influence in the first year. An example of this replacement process for $w_0=0.5$ can be seen in Figure \ref{fig:modification}. Figure \ref{fig:ar1_kdes} shows how this modification shifts the global distribution of autocorrelation to higher values. 

\end{doublespacing}
\newpage
\bibliography{references,main}
\newpage
\section*{Extended figures}

\setcounter{equation}{0}
\setcounter{figure}{0}
\setcounter{table}{0}
\setcounter{page}{1}
\makeatletter
\renewcommand{\theequation}{S\arabic{equation}}
\renewcommand{\thefigure}{S\arabic{figure}}
\renewcommand{\thetable}{S\arabic{table}}
\renewcommand{\bibnumfmt}[1]{[S#1]}
\renewcommand{\citenumfont}[1]{S#1}

\begin{figure}[htb!]
    \centering
    \includegraphics[width=\textwidth]{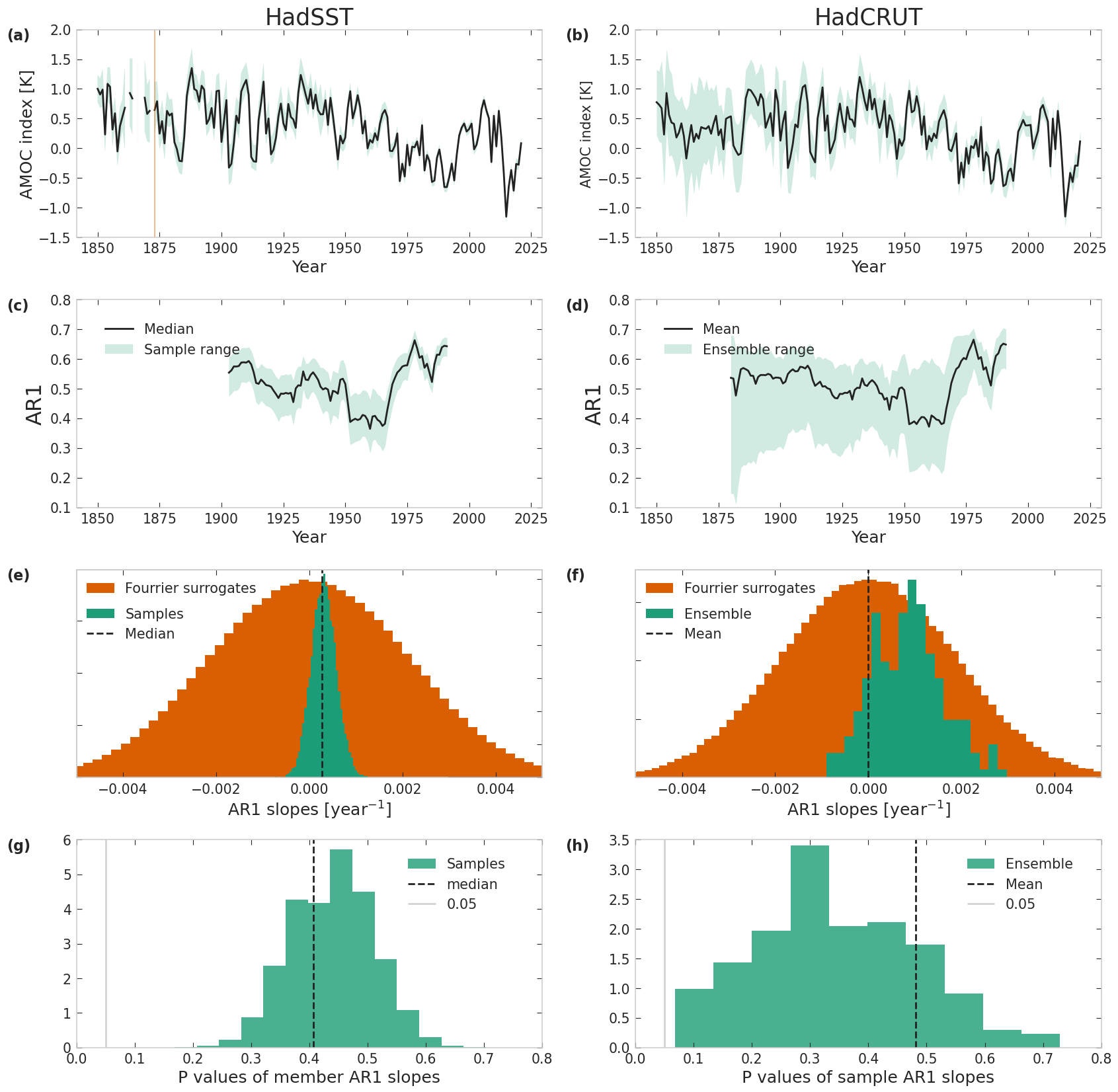}
    \caption{Same as Fig.~\ref{fig:indices} but for the AR1. The increase in AR1 is not significant for either HadSST and HadCRUT, although there is a large spread of p-values and for HadCRUT some ensemble members do show a significant increase.}
    \label{fig:ar1_indices}
\end{figure}
\begin{figure}[htb!]
    \centering
    \includegraphics[width=\textwidth]{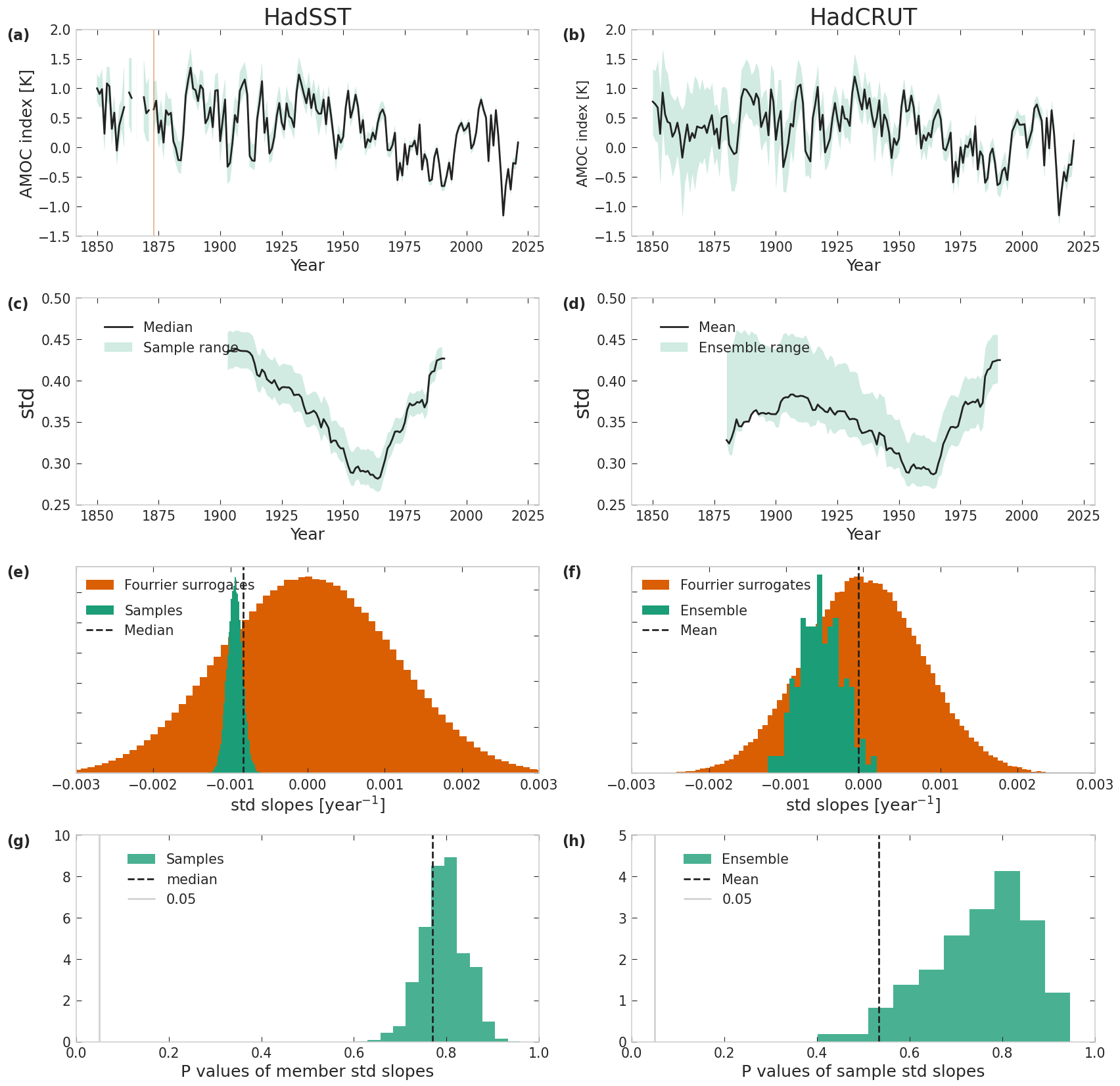}
    \caption{Same as Fig.~\ref{fig:indices} but for the standard deviation (STD). The STD for the AMOC index in HadSST and HadCRUT shows a marked difference from that of HadISST (see Fig \ref{fig:SST_significance_std} and B21, Fig.~3e). The overall negative trends of variance are because of a decrease in variance from 1900 to 1960, present in both HadSST and HadCRUT, but not in HadISST. This is likely due to the changing number of observations in the subpolar gyre (see Fig \ref{fig:nobs}): as the number of observations increase the index is a mean of more values and the variance decreases.}
    \label{fig:std_indices}
\end{figure}
\begin{figure}[htb!]
    \centering
    \includegraphics[width=\textwidth]{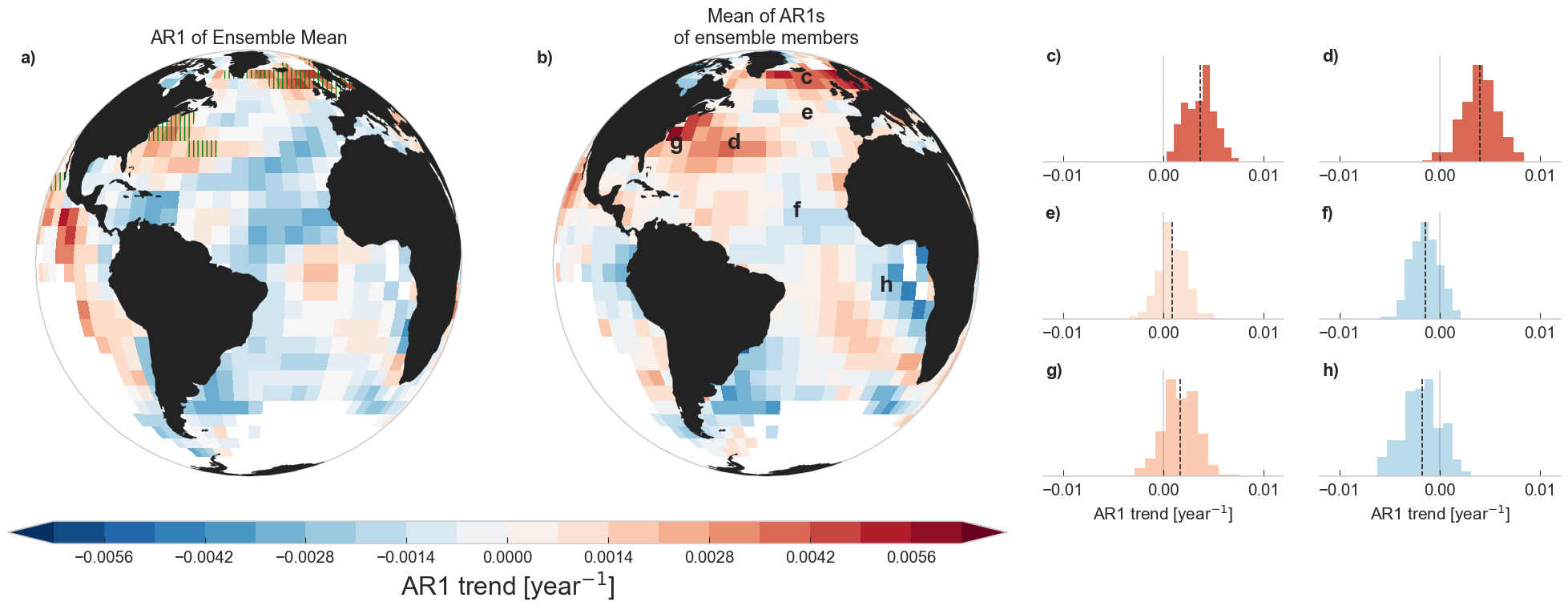}
    \caption{Same as Fig.~\ref{fig:hadcrut} but for the AR1 trends.}
    \label{fig:HadCRUT_ar1}
\end{figure}
\begin{figure}[htb!]
    \centering
    \includegraphics[width=\textwidth]{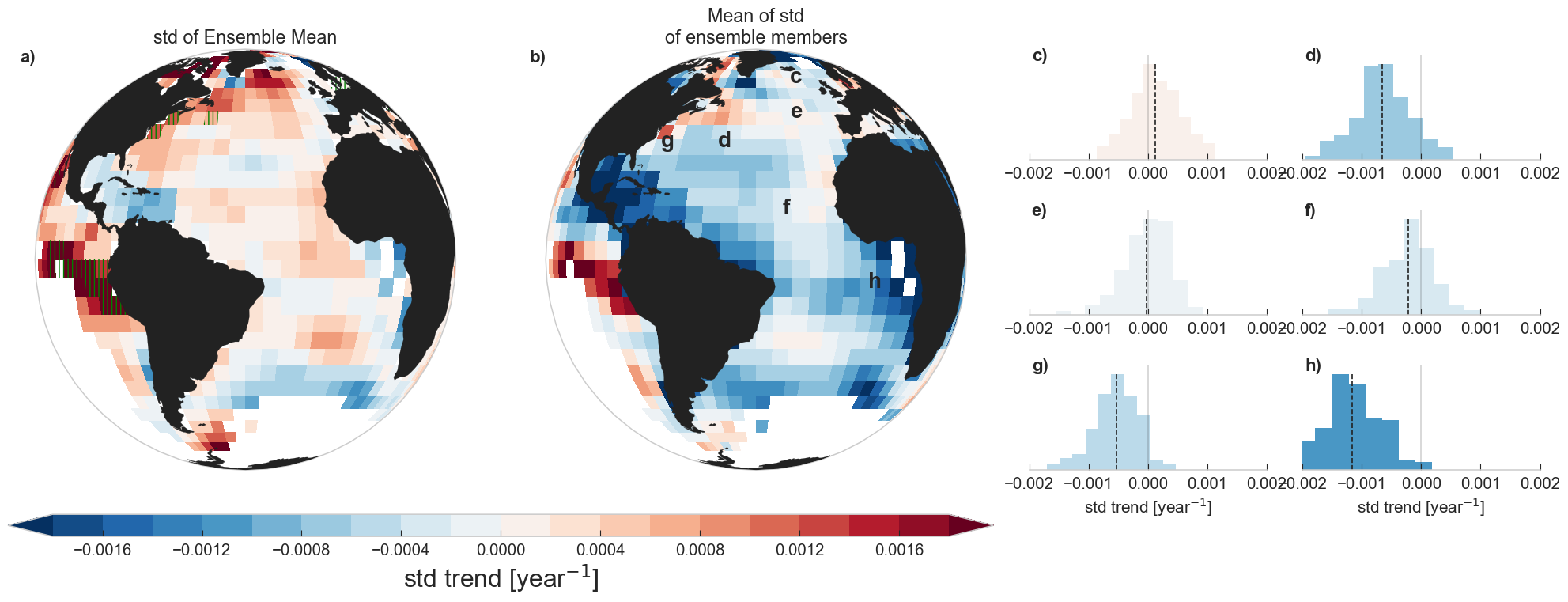}
    \caption{Same as Fig.~\ref{fig:hadcrut} but for the standard deviation (STD) of linear trends. The STD of the ensemble members is different from the STD of the ensemble mean because taking the mean removes the high variability at early times that causes the negative trend in the STDs seen in plots c-e (see also black line vs green range in Fig.~\ref{fig:std_indices}d,f and h).}
    \label{fig:HadCRUT_std}
\end{figure}

\begin{figure}[htb!]
    \includegraphics[width=\textwidth]{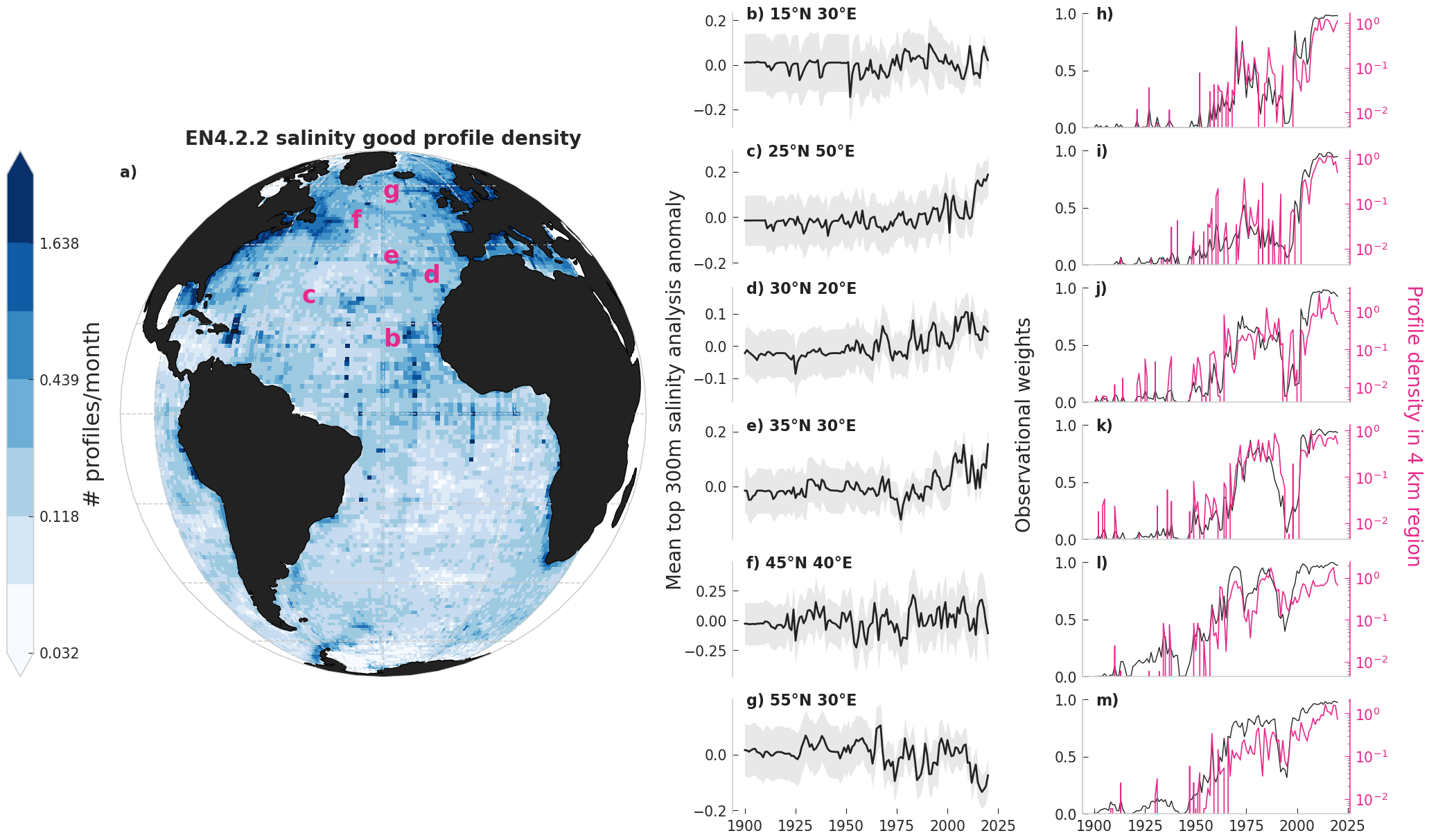}
    \caption{a. Number of good salinity profiles per month in the Atlantic Ocean for the EN4.2.2 dataset (note the logarithmic scale). b-g. Mean annual salinity in the upper 300m (black) with its corresponding analysis uncertainty range (grey shading) for six locations in the North Atlantic. h-m. Mean annual observational uncertainty (black) and the density of good salinity profiles in the surrounding 4$\degree$ region (pink) for the same six locations.}
    \label{fig:nprof}
\end{figure}
\begin{figure}[htb!]
    \centering
    \includegraphics[width=\textwidth]{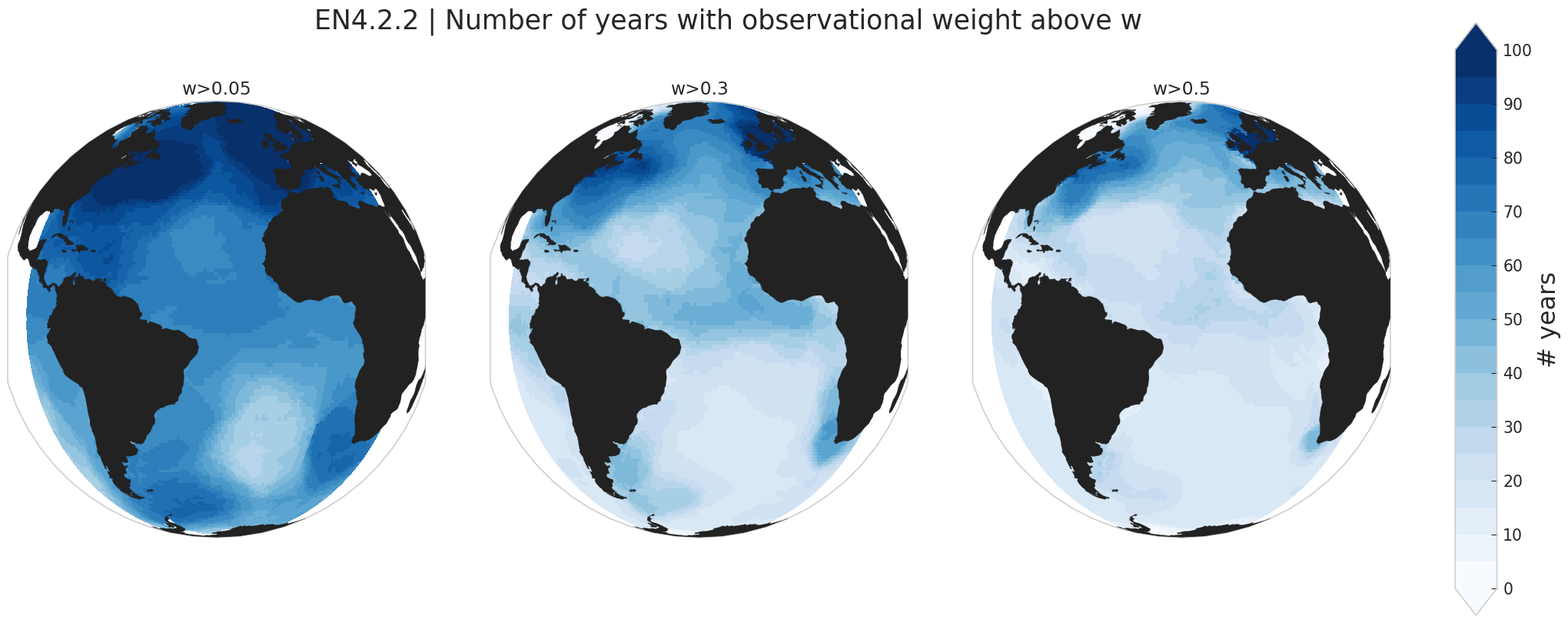}
    \caption{Number of years in the EN4.2.2 data with top 300m mean observational weight above w = 0.05 (a), 0.3 (b) and 0.5 (c). As in other figures only the region between 90$^\circ$W and 30$^\circ$E is shown.}
    \label{fig:w_years}
\end{figure}
\begin{figure}[htb!]
    \centering
    \includegraphics[width=\textwidth]{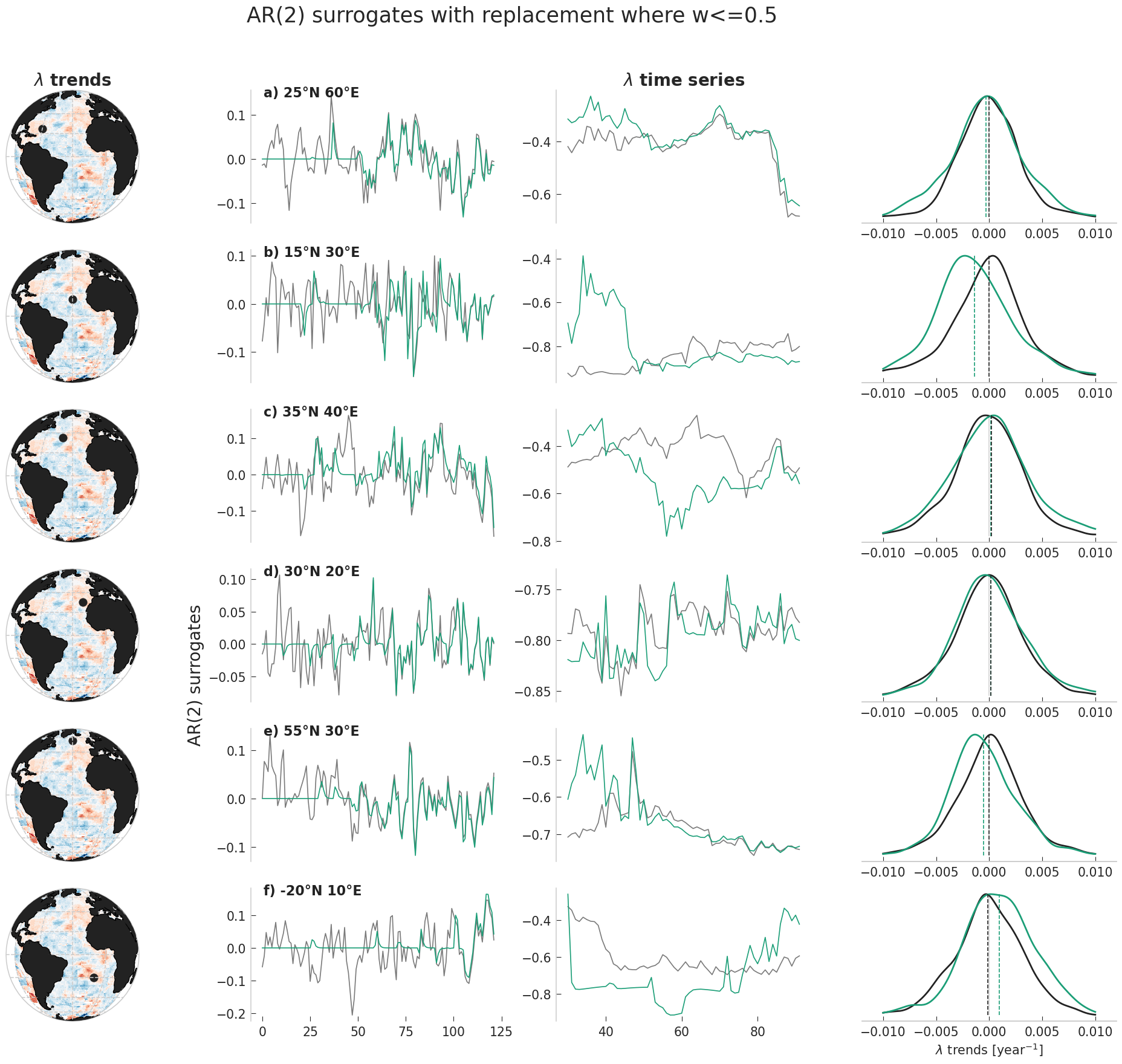}
    \caption{Modification of AR2 surrogates of salinity time series (see Methods). First column: mean of distribution of modified surrogate $\lambda$ trends. Second column: original AR2 surrogate (black) and modified AR2 surrogates where $w\leq0.5$ (turquoise). Third column: $\lambda$ time series of the time series shown in the second column. Fourth column: distribution of 1000 $\lambda$ trends at the location, with a dashed line of the corresponding colour showing the mean. Note that the modification does not systematically bias the trends.}
    \label{fig:modification}
\end{figure}
\begin{figure}[htb!]
    \centering
    \includegraphics[width=0.8\textwidth]{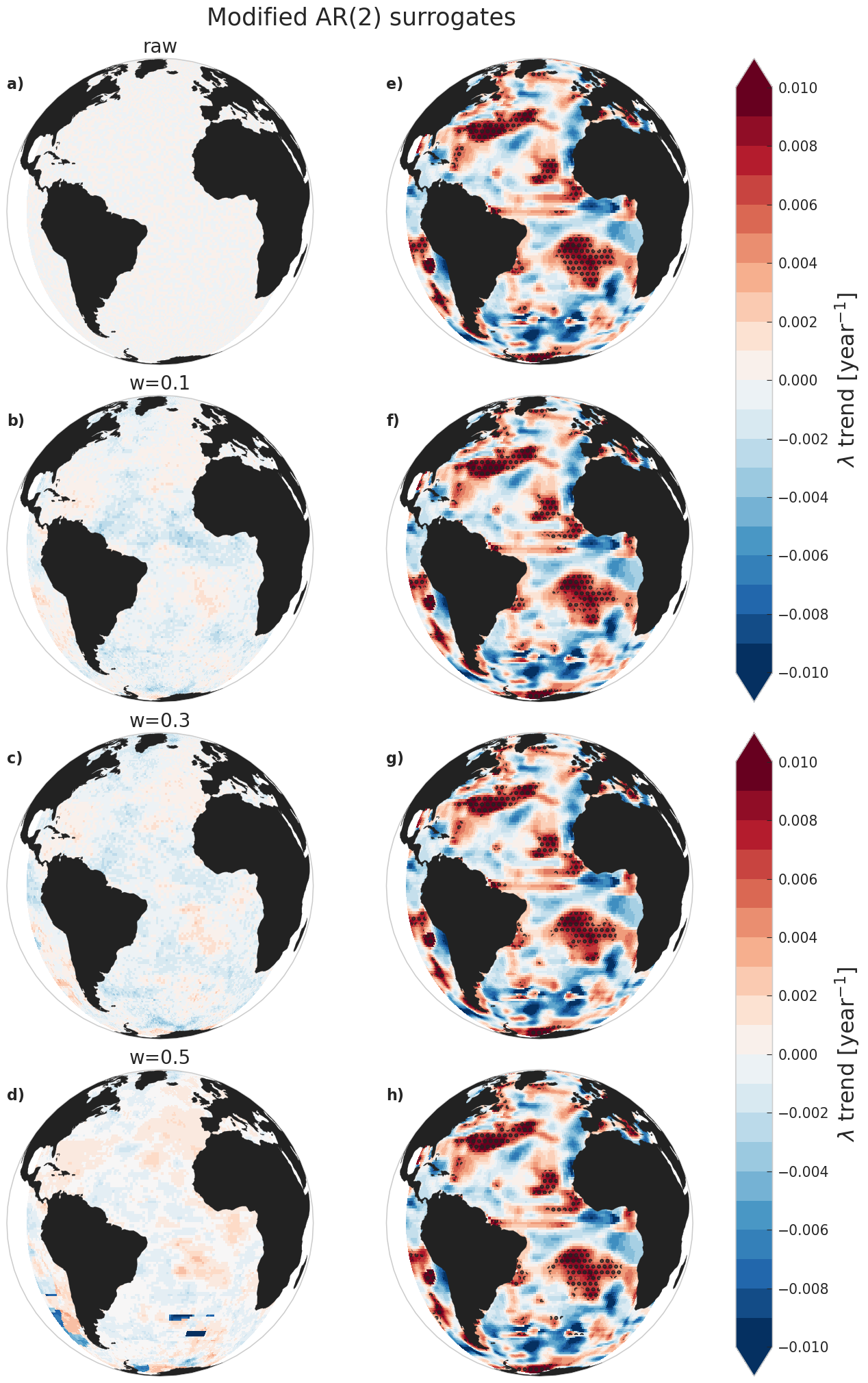}
    \caption{a. Mean of distribution of surrogate $\lambda$ trends for raw AR2 surrogates of salinity time series, and modified surrogates with observational weight limit b. w = 0.1, c. 0.3 and d. 0.5.
    The areas where the analysis trends are significant given each of the surrogate distributions is shown in the second row on top of the $\lambda$ trend of the analysis data (e-h). Note the differences in scale between the induced negative and positive biases (upper row) and the actually inferred trends (lower row). As in other figures only the region between 90$^\circ$W and 30$^\circ$E is shown.}
    \label{fig:elim_options}
\end{figure}

\begin{figure}[htb!]
    \centering
    \includegraphics[width=\textwidth]{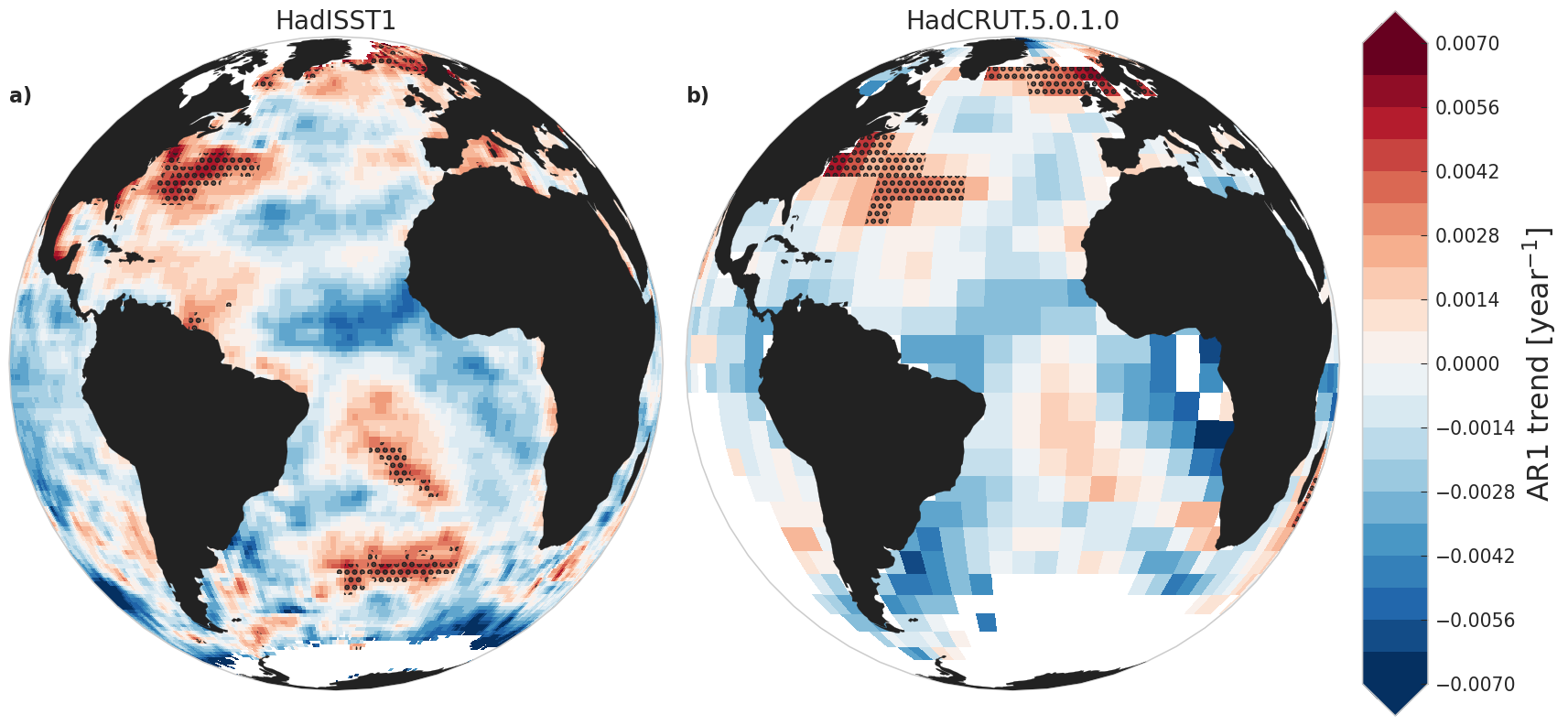}
    \caption{Same as Fig.~\ref{fig:sstsig} but for the AR1 trends. }
    \label{fig:SST_significance_ar1}
\end{figure}
\begin{figure}[htb!]
    \centering
    \includegraphics[width=\textwidth]{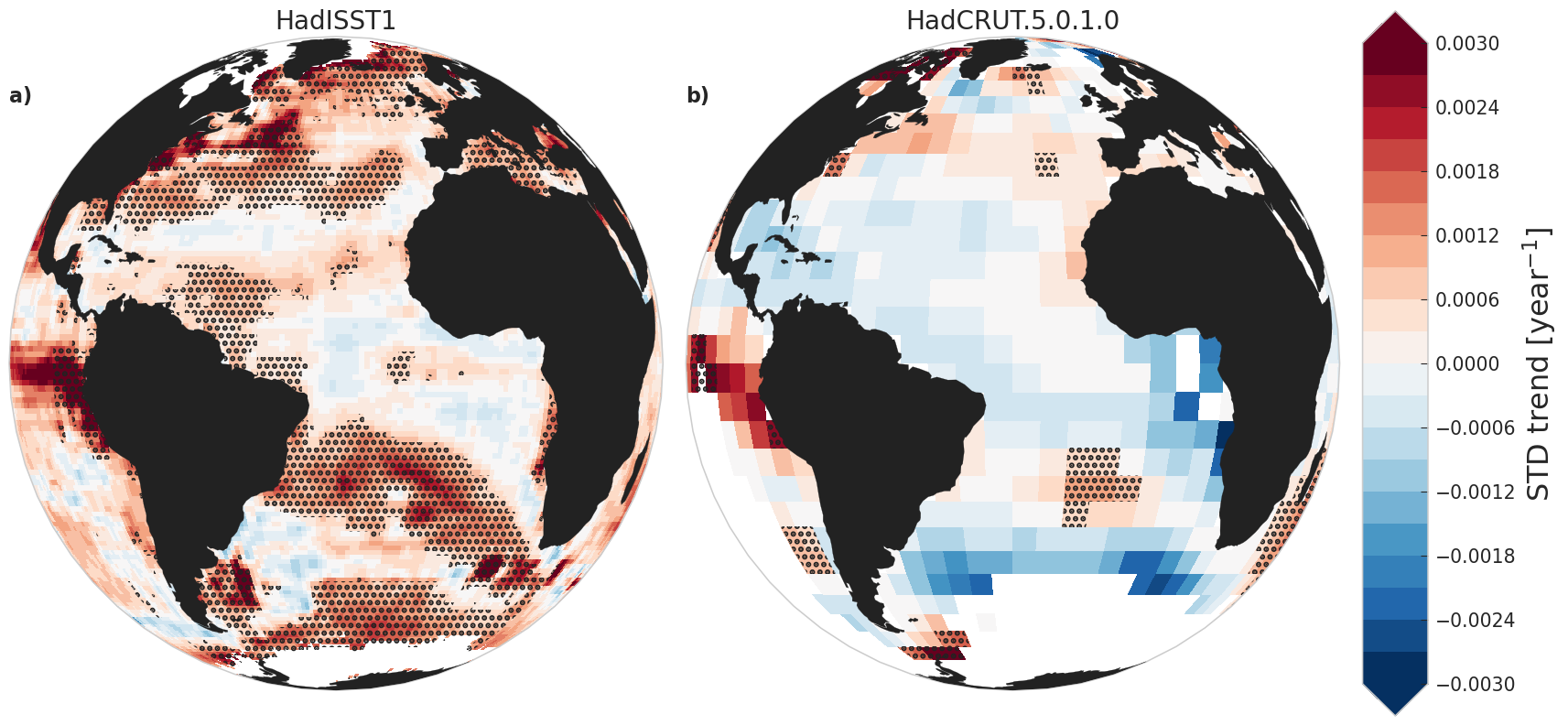}
    \caption{Same as Fig.~\ref{fig:sstsig} but for the STD trends. The significant difference between the trends for HadISST and HadCRUT is likely due to dataset properties, which have a large influence on the variance.
    }
    \label{fig:SST_significance_std}
\end{figure}
\begin{figure}[htb!]
    \centering
    \includegraphics[width=\textwidth]{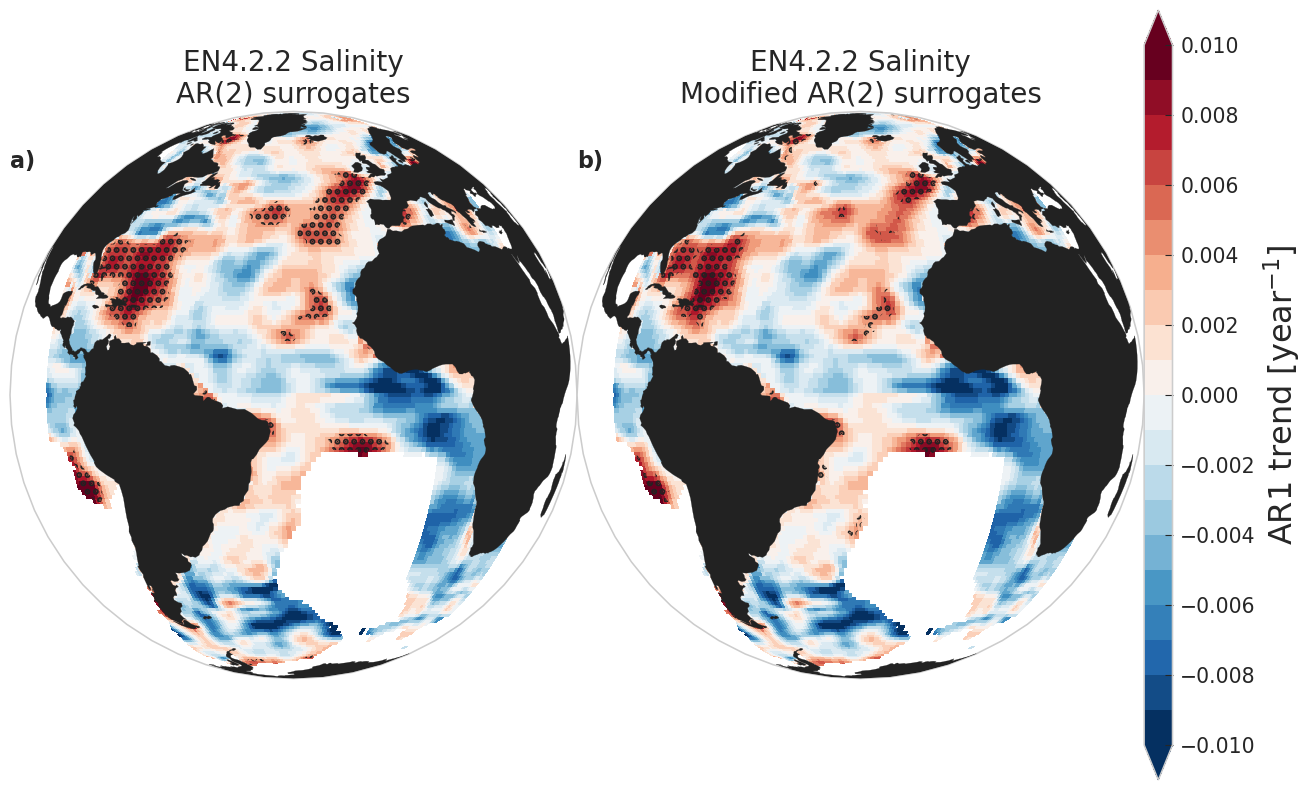}
    \caption{Similar to Fig.~\ref{fig:salsig} for the AR1, but showing the significance regions both for the raw AR(2) surrogates (a) and the modified surrogates with $w=0.5$ (b). The general location of positive and negative trends is similar to that for $\lambda$, but the regions of significance are different. Note that in this case the significance regions are slightly reduced when using the modified surrogates, as the modification causes a larger false increase in AR(2) than it does in $\lambda$. Nevertheless, the overall pattern of regions with significant increases remains qualitatively the same.
    }
    \label{fig:salinity_significance_ar1}
\end{figure}

\begin{figure}[htb!]
    \centering
    \includegraphics[width=\textwidth]{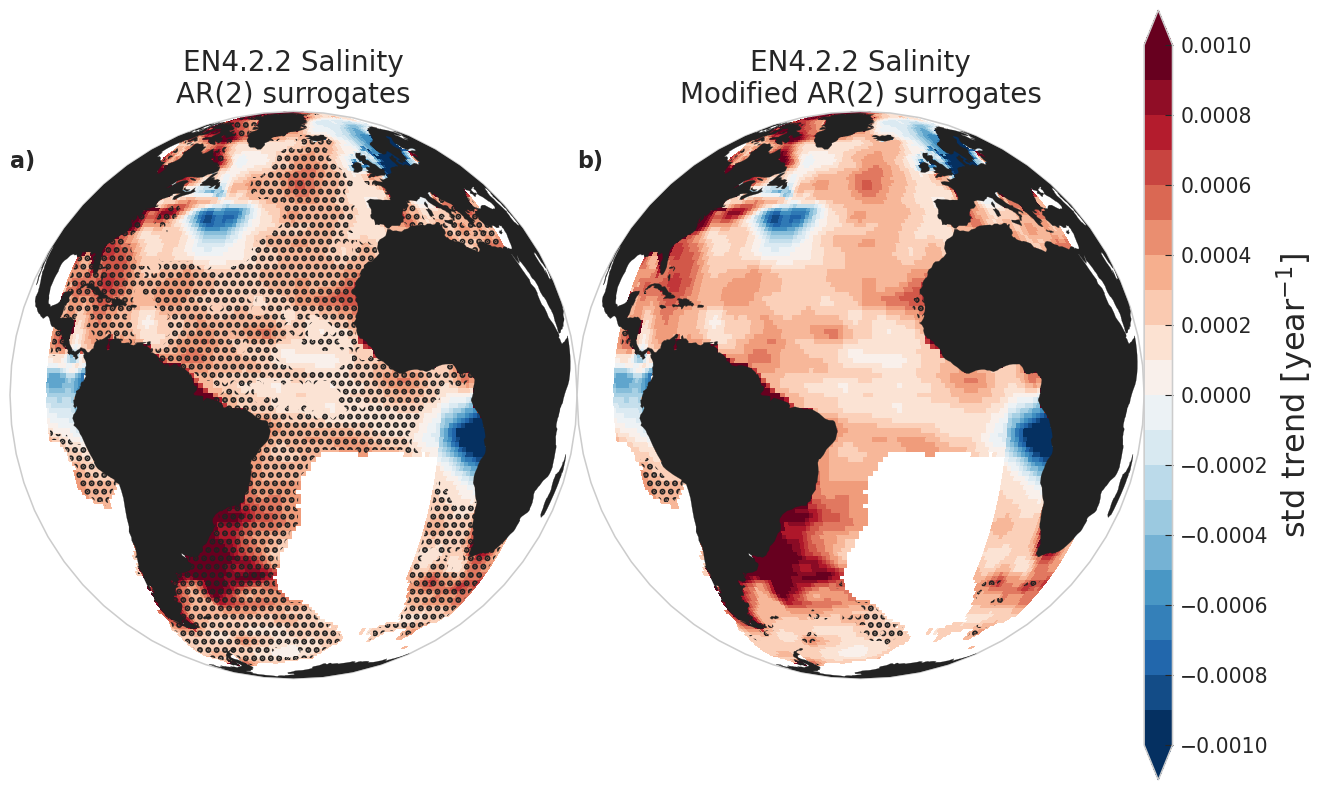}
    \caption{Same as Fig.~\ref{fig:salinity_significance_ar1} but for the STD. Although almost the whole of the Atlantic shows an increasing variance, using modified surrogates removes all of the significant regions because the modification mimics the false variance increase that the EN4 analysis method causes. The areas of strong negative trend close to the North American and South African coasts are also due to dataset irregularities (a "spike" in the early 1900, probably due to a faulty data point). Note that these results do not rule out the existence of increasing variance trends in the true SST values, as those would be masked by the analysis method that causes the false trends.}
    \label{fig:salinity_significance_std}
\end{figure}

\begin{figure}[htb!]
    \centering
    \includegraphics[width=\textwidth]{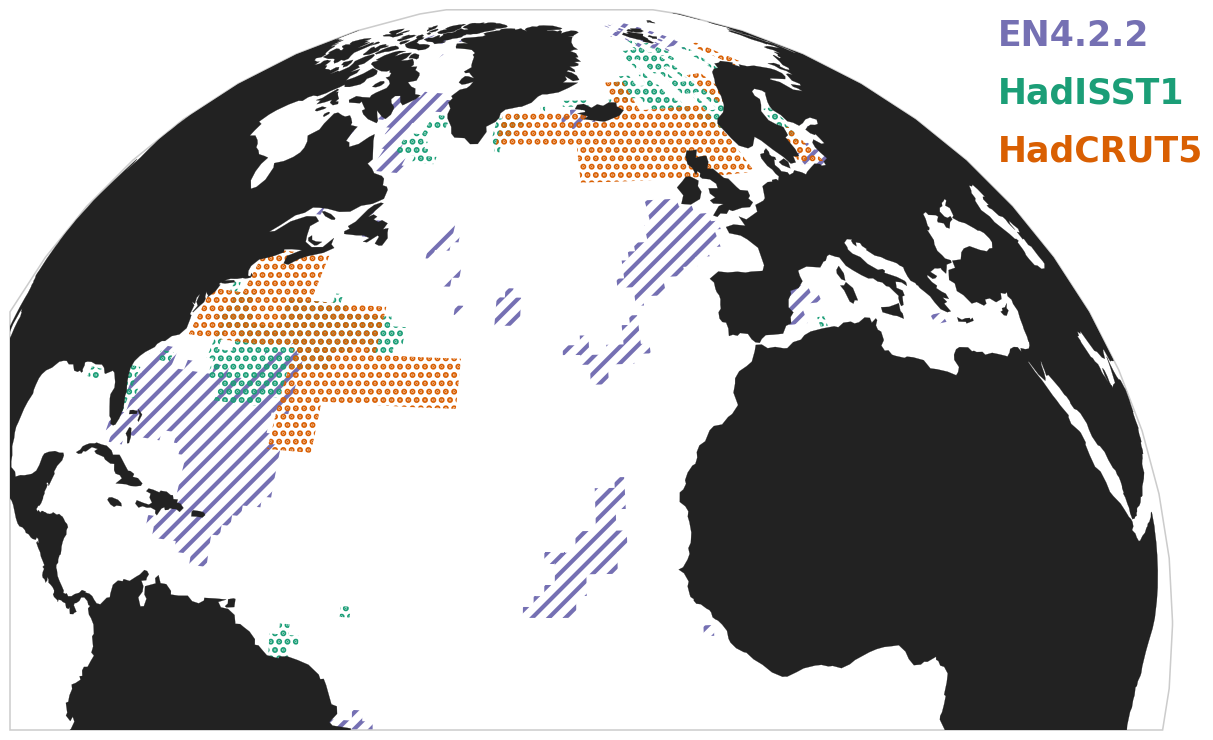}
    \caption{Same as Fig.~\ref{fig:sig_regions} but for the AR1. Similarly to $\lambda$, the significance regions for the three datasets are roughly in the same area. The regions in the Irminger, Greenland and Iceland seas are similar to those found for $\lambda$. However the significant regions in lower latitudes are the northern Gulf Stream and its extension into the Atlantic Ocean. }
    \label{fig:ar1_sig_regions}
\end{figure}
\begin{figure}[htb!]
    \centering
    \includegraphics[width=\textwidth]{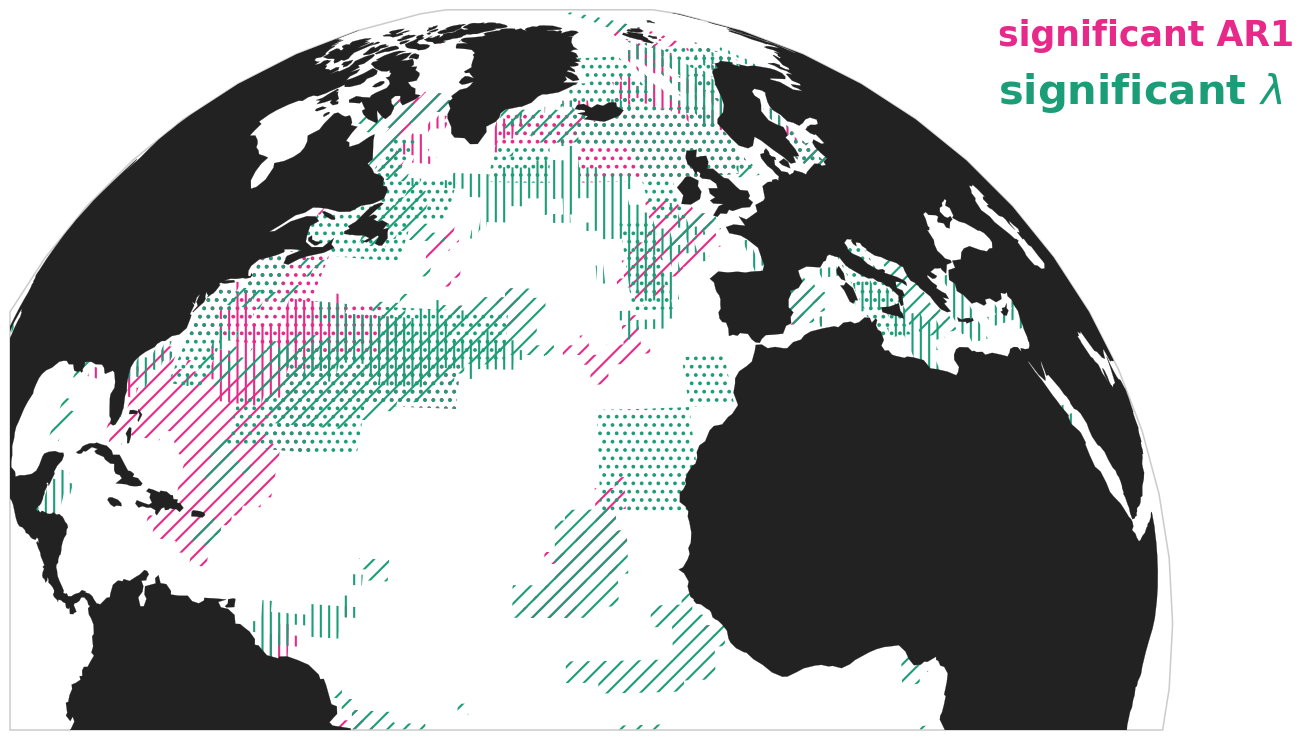}
    \caption{Same as Fig.~\ref{fig:sig_regions} but for both the AR1 and $\lambda$ trends. All the significant AR1 ($\lambda$) regions are shown in pink (turquoise). The three datasets are distinguished by pattern: dots, vertical lines and diagonal lines for HadCRUT, HadISST and EN4, respectively. The difference between the two indicators is due to the calculation method of $\lambda$, which accounts for autocorrelated residual noise. Together the regions for both indicators trace the currents that are part of the AMOC in the North Atlantic: from the Gulf Stream along the North Atlantic Current and into the Greenland, Iceland, Irminger and Labrador seas. }
    \label{fig:both_sig_regions}
\end{figure}
\begin{figure}[htb!]
    \centering
    \includegraphics[width=\textwidth]{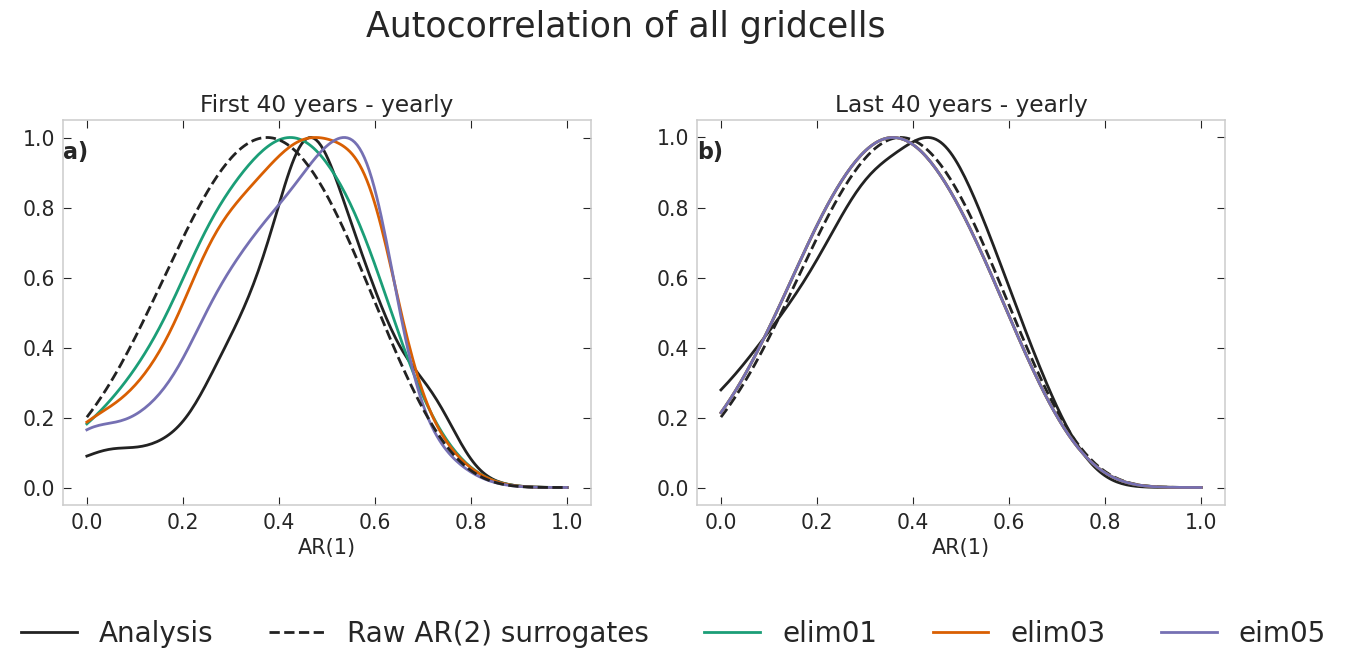}
    \caption{To show the difference between the first and last 40 years of salinity analysis data and the effect that modification has on the surrogates, we plot the distribution of the lag-one autocorrelation values of all Atlantic grid cells in the EN4.2.2 dataset in the first (a) and last (b) 40 years of data. The analysis autocorrelation (solid black) changes from a sharply peaked distribution in the first 40 years to a wider one in the last 40 years. The raw AR(2) surrogates (dashed black) match the last 40 years, but not the first 40. The modified surrogates with weight limits of 0.1 (turqouise), 0.3 (orange) and 0.5 (purple) have a more similar distribution to the analysis in the first 40 years, and are unchanged from the raw surrogates in the last 40 years, as required.}
    \label{fig:ar1_kdes}
\end{figure}
\end{document}